\documentclass[twocolumn]{aastex631}

\usepackage[utf8]{inputenc}
\usepackage{xcolor}
\usepackage{amsmath}

\begin{document}

\title{Detectability of Self-Lensing Flares of White Dwarfs with Compact Companions}

\author[0000-0002-7501-5579X]{Guy Nir}
\author[0000-0002-7777-216X]{Joshua S.~Bloom}
\affiliation{Department of Astronomy, University of California,
Berkeley, CA 94720-3411, USA}
\affiliation{Lawrence Berkeley National Laboratory, 1 Cyclotron Road,
MS 50B-4206, Berkeley, CA 94720, USA}

\date{\today}

\begin{abstract}
    Binaries containing compact objects, if viewed close to edge on, 
    can produce periodic brightening events under certain conditions
    on the masses, radii, and binary separation. 
    Such flares are caused by one object gravitational lensing the other, 
    in what is known as \emph{self-lensing flares}. 
    We present a simulation tool that efficiently reproduces
    the main features of self-lensing flares and 
    facilitates a detection sensitivity analysis for various sky surveys.
    We estimate the detection prospects for a handful of representative
    surveys when searching for systems of either two white dwarfs, 
    or a white dwarf with other compact objects, 
    i.e., neutron stars and black holes. 
    We find only a marginal ability to detect such systems in existing surveys.
    However, we estimate many such systems could be detectable by 
    surveys in the near future, including the Vera Rubin observatory. 
    We provide a quantitative analysis of the detectability 
    of double-compact object self-lensing flares across 
    the landscape of system parameters, 
    and a qualitative discussion of 
    survey and followup approaches 
    to distinguish such flares from confounding events, 
    such as stellar flares, 
    satellite glints, and cosmic rays. 
    We estimate 0.3, 3 and 247 double white dwarf systems 
    could be detected by TESS, ZTF, and LSST, respectively. 
    A similar number of systems with a neutron star or black hole companion
    could be detected, but we caution that the 
    number densities of such binaries is model dependent and so our detection estimates. 
    Such binaries can be used to constrain models of the end states of binary evolution.
\end{abstract}

\section{Introduction}\label{sec: introduction}

The general relativistic deflection of light passing near a massive object can lead to
apparent astrometric and brightness changes, so-called gravitational lensing
 \citep{microlensing_stars_Einstein_1936,microlensing_Refsdal_1964}.  
For background stars lensed by foreground compact objects in our galaxy, 
the multiple images of the star created during gravitational lensing are usually too small to resolve, 
a situation referred to as \emph{microlensing} \citep{microlensing_MACHOs_Paczynski_1986}. 
However, in many real-world cases, a brightness change with time as well as an
astrometric deflection, can be observed.
Such measurements can be
used to find otherwise invisible compact objects
such as black holes, or other dark massive objects
\citep[e.g.,][]{microlensing_black_holes_MACHO_Bennett_2002, microlensing_astrometric_Rybicki_2018, microlensing_OGLE_galactic_bulge_Mroz_2019, microlensing_black_hole_Lam_2022}. 

Self Lensing (SL) occurs when one component in a binary system
magnifies the light from the other component. 
In this case the magnification amplitude is typically smaller 
than for lensing by unrelated, distant bodies, 
and the time scales of the flares are much shorter:
minutes or hours as opposed to days or weeks for 
the more typical background-foreground lensing in the Galaxy
\citep{self_lensing_degenerate_binaries_Maeder_1973}. 
Such systems can provide much information 
on the physical properties of the binary:
the orbit must be nearly edge-on for the lensing effect to be detectable, 
and the duration and amplitude of the flare provide constraints
on the mass of the lensing body. 
The lensing flare is periodic, 
so repeat measurements can lead to determination of the orbital period. 
Combined with radial velocity measurements, 
the system parameters can be measured to high precision. 
To date, most binaries with compact objects, 
specifically neutron stars (NSs) or black holes (BHs), 
have been detected by the \hbox{X-ray} flux from accretion 
of gas from a close companion. 
Such \hbox{X-ray} binaries can only provide information
on very close systems, 
that are undergoing active mass transfer. 
On the other hand, 
self-lensing systems can help 
shed light on a population of non-interacting, 
wide binaries with compact objects
\citep{self_lensing_Beskin_Tuntsov_2002}. 

The typical effective size of the gravitational effect, 
the \emph{Einstein radius}, 
is between $10^3$ and $10^5$\,km, 
depending on the orbital separation and mass of the lens. 
For orbits smaller than an AU, 
the lens is much smaller than a solar radius, 
so flares from main sequence stars 
with compact companions are hard to detect
\citep{self_lensing_compact_binaries_Qin_1997, self_lensing_Beskin_Tuntsov_2002, self_lensing_main_sequence_sources_Rahvar_2011}. 
\cite{self_lensing_MS_NS_BH_Wiktorowicz_2021} estimate hundreds
of SL flares from MS sources (with NS or BH companions) 
can be detected in all-sky surveys. 
However, many of the flares they discuss 
are either very faint (with an amplitude less than 1\% of the star's light), 
or have a single detection during the life of the survey,
so they would be very hard to distinguish 
from stellar flares \citep{stellar_flare_foreground_Kulkarni_Rau_2006} 
or satellite glints \citep{satellite_glints_Corbett_2020,satellite_glints_Nir_2021}. 

On the other hand, 
if the source is a luminous compact object, 
i.e., a white dwarf (WD), 
then the flares can become substantial, 
with magnifications of up to a few times 
the quiescent flux, 
even for relatively short period orbits. 
In some cases, \hbox{X-rays} from accretion 
on to a NS or BH can also be magnified periodically 
by a compact lens
\citep{self_lensing_x_rays_DOrazio_2020}.

WDs are therefore favorable sources 
for detection of SL flares, 
albeit with a smaller total number of observable systems
compared to MS sources. 
In cases where the compact object is a NS or BH, 
the temporal behavior and brightness of the flare is determined 
by the size of the Einstein radius and the 
finite size of the source. 
For a WD lens occulting a WD source, 
the physical size of the lens must also 
be taken into account. 
In such cases the shape of the lightcurve 
would be determined by a competition 
between the lens magnification and the 
physical occultation by the lens
\citep[e.g.,][]{self_lensing_planet_transits_new_observable_Kasuya_2011, self_lensing_degeneracy_occultations_Han_2016}.

Binaries with wide orbits (on the order of AUs)
typically produce brighter flares. 
Close orbits (with semimajor-axis $a<10^{-2}$\,AU) 
would produce faint flares 
(on the order of few percent increase in flux), 
with minute-scale durations. 
While detection of such short, 
faint flares is challenging, 
such systems would have short periods 
(on scales of hours), 
and so would be easier to follow up and verify. 

Self-lensing binaries are rare and hard to detect. 
The system must be close to edge-on, 
and the flare occurs only within a small 
fraction of the orbit (low duty cycle). 
Double White Dwarf (DWD) systems are fairly common, 
where the fraction of WDs in a binary is 0.1--1, 
depending on the separation \citep{white_dwarf_binary_fraction_Toonen_2017, white_dwarfs_binary_separation_Maoz_2018}. 
However, the probability of finding a DWD with a sufficiently large
inclination angle (i.e., almost edge on) significantly reduces
the number of observable systems, especially when considering 
wider orbits with $a= 10^{-2}$--$10^{-1}$\,AU. 

NS and BH companions to WDs 
are expected to be even rarer, given the steep mass 
distribution of stars \citep{population_black_holes_Wiktorowicz_2019}. 
In this case, however, the higher mass and smaller
physical size of the lensing object may make them detectable 
at closer orbits, which means more systems
will be sufficiently inclined to have visible flares. 
The shorter periods would also give such systems a higher
duty cycle and shorter flare repeat time, 
making them easier to detect and follow up. 
It should be noted that WDs in binaries with pulsars
(fast spinning NSs) have been identified 
\cite{white_dwarfs_orbiting_pulsars_Istrate_2014, white_dwarf_with_massive_companion_2023}. 

\cite{self_lensing_Beskin_Tuntsov_2002} estimated up to 22 WD-WD pairs
and up to 16 WD-BH pairs can be detected using the same telescope as was used by 
the Sloan Digital Sky Survey (SDSS; \citealt{sloan_digital_sky_survey_technical_overview_York_2000}).  
\cite{black_hole_binaries_TESS_Masuda_2019} and 
\cite{black_hole_binaries_photometric_Chawla_2023} discussed the prospects 
of finding black holes using the self-lensing signal, 
among other indicators such as ellipsoidal variation and relativistic beaming. 

All-sky surveys increasingly push the boundaries
of sky coverage, cadence, and photometric precision. 
The Zwicky Transient Facility \citep[ZTF;][]{Zwicky_transient_facility_Bellm_Kulkarni_2019}, 
The All-Sky Automated Survey for Supernovae \citep[ASAS-SN;][]{ASAS_SN_Shappee_2014}, 
and the future Vera Rubin Legacy Survey of Space and Time \citep[LSST;][]{large_synoptic_survey_telescope_Ivezic_2007, large_synoptic_survey_telescope_Ivezic_2019}
have large fields of view (multi-square degree) and cadence
from few-per-night to once-per-few-nights. 
The Dark Energy Camera (DECam; \citealt{decam_dark_energy_camera_DePy_2008}) conducts multiple surveys, 
including a Deep Drilling Fields survey that takes multiple images
of the same fields over multiple, consecutive nights with LSST-like depth
\citep{decam_deep_drilling_field_DDF_Graham_2023}. 
Space telescopes such as the Transiting Exoplanet Survey Satellite \citep[TESS;][]{transiting_exoplanet_survey_satellite_Ricker_2014},
and the future Cubesats for Rapid Infrared and Optical Surveys (CuRIOS; \citealt{curios_slides_spie_Lu_Gulick_2022}) 
will observe large fractions of the sky continuously 
at high cadence and high photometric precision, 
while similar efforts to provide continuous imaging 
of large fields of view every few seconds 
are also planned and ongoing, 
e.g., the Weizmann Fast Astronomical Survey Telescope \citep[W-FAST;][]{Weizmann_fast_astronomical_survey_telescope_Nir_2021}
the Large Array Survey Telescopes \citep[LAST;][]{small_telescopes_LAST_Ofek_Ben_Ami_2020}, 
the Evryscope \citep{evryscope_design_Ratzloff_2019} 
and the Argus Optical Array \citep{argus_optical_array_Law_2021}. 

Since each such survey has dramatically different
cadence, field of view, depth and precision, 
we present in this work a detailed analysis 
of the detection probability to find self-lensing flares
from different types of binaries, 
given the properties and observing strategy of each survey. 
Using simple models for the distributions of 
mass, temperature, and binary separations 
of WDs with other WDs or more massive compact objects, 
we estimate the expected number of detections 
of self-lensing systems in various surveys.
The simulations described in this work can be repeated 
(for the given surveys, or by defining other survey parameters)
using an open source software package.\footnote{
\url{https://github.com/guynir42/self_lens}
}

We discuss the details of the simulation tools used
to produce the self-lensing flares in \S\ref{sec: simulator}. 
We present the physical properties of self-lensing systems in \S\ref{sec: physical properties}
and survey the landscape of the parameter space of such systems in \S\ref{sec: parameter space}. 
We discuss the interplay of astronomical surveys' properties
in regards to the chances of detection of self-lensing flares in \S\ref{sec: survey strategies}. 
We present the results of the simulations, including each survey's ability
to detect self-lensing flares in \S\ref{sec: results}. 
We discuss these results, and compare them to other cases of self-lensing systems
and eclipsing systems in \S\ref{sec: discussion}. 
We conclude in \S\ref{sec: conclusions}.

\section{Self-lensing simulator}\label{sec: simulator}

Light is deflected around massive objects 
and can be focused by a point mass in such a way 
that amplifies the amount of light reaching an observer. 
A point source close to such a massive object (``lens"), 
would be seen as two images. 
The positions of each image are given by 
the lens equation \citep{microlensing_MACHOs_Paczynski_1986, self_lensing_Beskin_Tuntsov_2002}:
\begin{equation}\label{eq: lens equation}
    u = d - \frac{1}{d},
\end{equation}
where $u$ and $d$ are the normalized distances 
(on the plane of the sky at the distance to the lens) 
from the center of the lens to the 
center of the source, 
and from the center of the source to the center of its images, 
respectively.
These two sizes are both given in units
of the Einstein radius: 
\begin{equation}\label{eq: einstein radius}
    R_E = \sqrt{\frac{4GM (D_S - D_L) D_L}{c^2 D_S}} \approx \sqrt{\frac{4GMa}{c^2}} 
\end{equation}
where $D_S$ and $D_L$ are the distances to the source and lens, respectively, 
the mass $M$ is of the lens, and $G$ and $c$ are the usual 
gravitational constant and speed of light. 
The approximation above holds 
when the lens is orbiting the source at a distance $a\ll D_S\sim D_L$. 
For $M$ in units of a Solar mass, and $a$ in AU, 
the Einstein radius is $R_E\approx 1.5\times 10^4 \sqrt{M a}$\,km.
We will assume a circular orbit in all calculations, 
so $a$ can also be referred to as the semimajor axis. 

If the lens and source are perfectly aligned, 
$u\to 0$ and we get a ring around the source position, 
with a radius given by the Einstein radius, 
from Equation~\ref{eq: einstein radius}. 
If the lens is not perfectly aligned 
we get two images with positions that
can be found by solving Equation~\ref{eq: lens equation}:
\begin{equation}\label{eq: image positions}
    d_{1,2} = \frac{1}{2} \left(u \pm \sqrt{u^2 + 4} \right).
\end{equation}

If the Einstein radius is much larger than the source, 
we can approximate the resulting lightcurves as point source events
(e.g., \citealt{microlensing_MACHOs_Paczynski_1986, microlensing_astrometric_Rybicki_2018}): 
\begin{equation}\label{eq: point source approximation}
    A = \frac{u^2 + 2}{u\sqrt{u^2 + 4}},
\end{equation}
where $A$ is the magnification of the source 
given the source-lens distance $u$. 

\begin{figure*}
    \centering
    
    \includegraphics[width=1.0\linewidth]{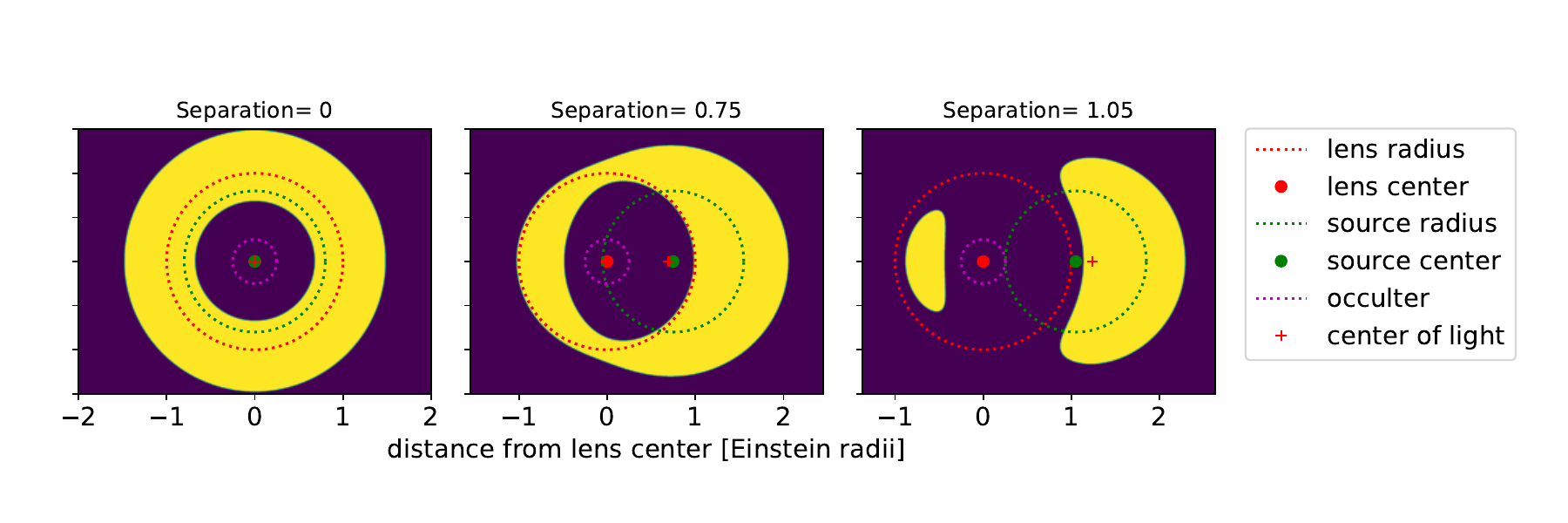}
    \caption{Examples of configurations of microlensing
             by a source with size 0.8 Einstein radius units. 
             Left: the lens and source are aligned.
             Middle: the lens is offset but still inside the source area.
             Right: the lens is outside the source. 
             The dark purple is outside the contours of the lensed image (no light) 
             and the bright yellow is inside the contours 
             (light imaged from the source surface). 
             }
    \label{fig: multiple geometries}
\end{figure*}

For a source of finite extent, 
Equation~\ref{eq: image positions} can be used to find the image
of the contour tracing the edges of the source. 
If the lens is inside the source area, 
we get two warped contours, one inside the other. 
The source is magnified and warped 
and has a hole inside 
where light is deflected away from the observer. 
If the lens center is outside the source region, 
the contours are separated, 
one inside the Einstein radius and one outside of it. 
We show three representative configurations in Figure~\ref{fig: multiple geometries}. 

The surface brightness of the source is preserved in the image(s), 
but the surface area is larger than the area of the source, 
such that the total amount of light is magnified when 
the lens passes near the source. 
Since in most cases the images are not symmetrical, 
there is also an offset of the center of light from 
the true source position. 
Thus, the effects of microlensing, 
even when the images cannot be resolved, 
can be detected either in photometric or astrometric measurements. 

We turn to estimating the magnification and offset
for each value of $r_s = R_s/R_E$, the normalized source size, 
and $u$ (the normalized separation of source and lens centers). 
For each geometry we draw a circular contour of size $r_S$, 
and map the points, using Equation~\ref{eq: image positions}, 
unto a digital binary image matrix. 
The result is an outline for two shapes. 
Inside the contours we set the binary image to one, 
and outside the contours to zero. 
For geometries where $u<r_s$ the shapes are one inside the other, 
so we subtract the small one from the big one 
(using a \texttt{xor} operation). 
In geometries where $u>r_s$, the images are separate and we 
calculate the results for each one separately. 

\begin{figure}
    \centering
    \includegraphics[width=1\linewidth]{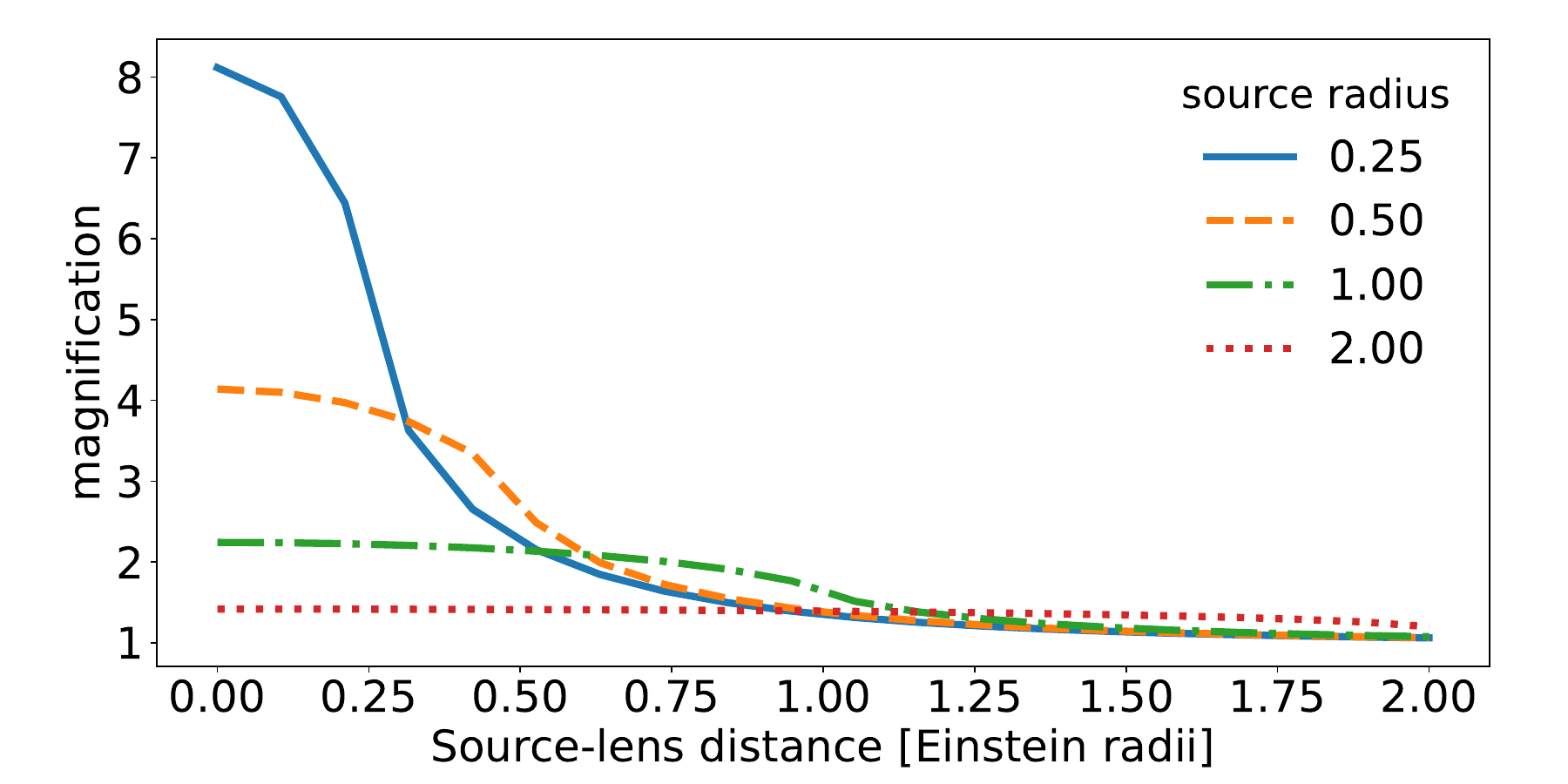}
    \caption{Magnification as a function of distance between lens and source centers (on the image plane). 
             Several sizes of the source are shown, in units of Einstein radius. 
             The smaller sources are more magnified, 
             but their magnification drops off faster with separation. 
    }
    \label{fig: example lcs sources}
\end{figure}

In these binary images, 
the relative area represents the total flux, 
while the first moment gives an estimate of the astrometric offset. 
If the binary map is given by $I$ 
and one of the image coordinates is given by $x$, 
then the flux $f$ and moment $m$
are given by
\begin{equation}\label{eq: flux and moment}
    f = \sum I, \quad m = \frac{\sum x I}{f}.
\end{equation}
Example magnification curves, 
as a function of distance between lens and source, 
for several source sizes, 
are shown in Figure~\ref{fig: example lcs sources}. 

\begin{figure}
    \centering
    \includegraphics[width=1\linewidth]{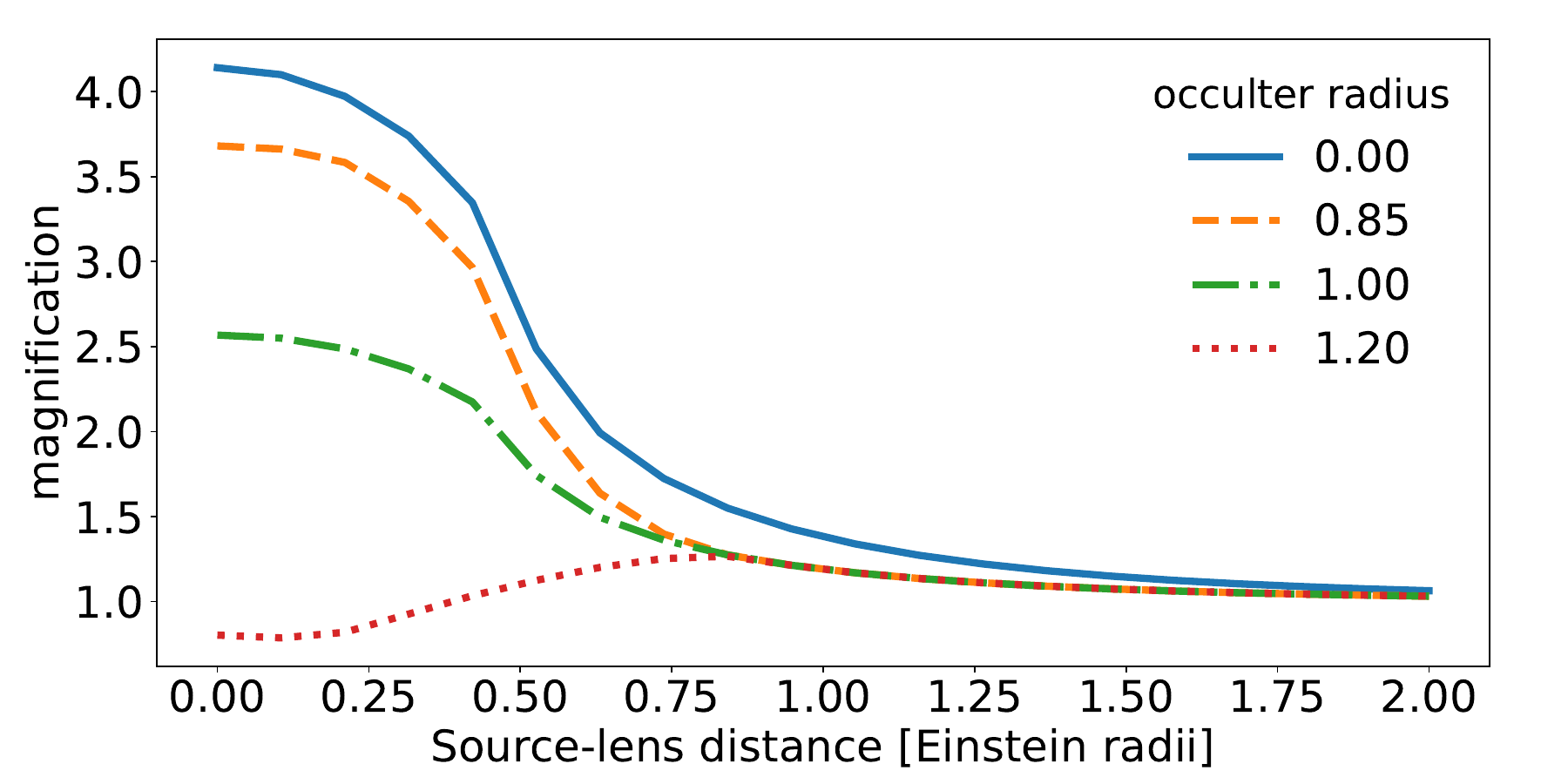}
    \caption{Magnification as a function of distance between lens and source centers (on the image plane). 
             Several sizes of the physical occulter are shown, in units of Einstein radius. 
             The source in this example has a size of 0.5\,$R_E$. 
             When the occulter is bigger than the source (0.85--1.00\,$R_E$)
             the lensing flare is somewhat weakened. 
             When the occulter is substantially larger than the source (1.2\,$R_E$)
             the flare is completely lost, and replaced with an eclipse. 
    }
    \label{fig: example lcs occulters}
\end{figure}

Self-lensing objects often have an Einstein radius 
on the order of a few thousands of km, 
similar to the size of planets or WDs. 
This leads to strong finite source effects
when the source is a WD or a main sequence star. 
When the lensing object is also a WD, 
its size could be close to, or even larger
than the source and the Einstein radius, 
leading to substantial eclipses at the same time
that the source is magnified by the lens
\citep{self_lensing_Beskin_Tuntsov_2002, self_lensing_degeneracy_occultations_Han_2016}. 
Adding this effect to our calculations is simple:
we subtract the region of the physical extent of the lens, 
given by the radius $r_L=R_L/R_E$, 
normalized to the Einstein radius, 
from the binary image $I$ before calculating 
the flux and offsets for each geometry. 
Thus, we can calculate the flux and offset
for any self-lensing geometry, 
characterized by the three parameters:
$u$, $r_S$ and $r_L$. 
Magnification curves for a constant source size
but for various occulter sizes are shown in 
Figure~\ref{fig: example lcs occulters}. 
In the case where the occulter is much larger 
than the source (1.2 vs.~0.5 Einstein radii)
the flare is completely suppressed 
and replaced by an eclipse. 

\begin{figure}
    \centering
    \includegraphics[width=1\linewidth]{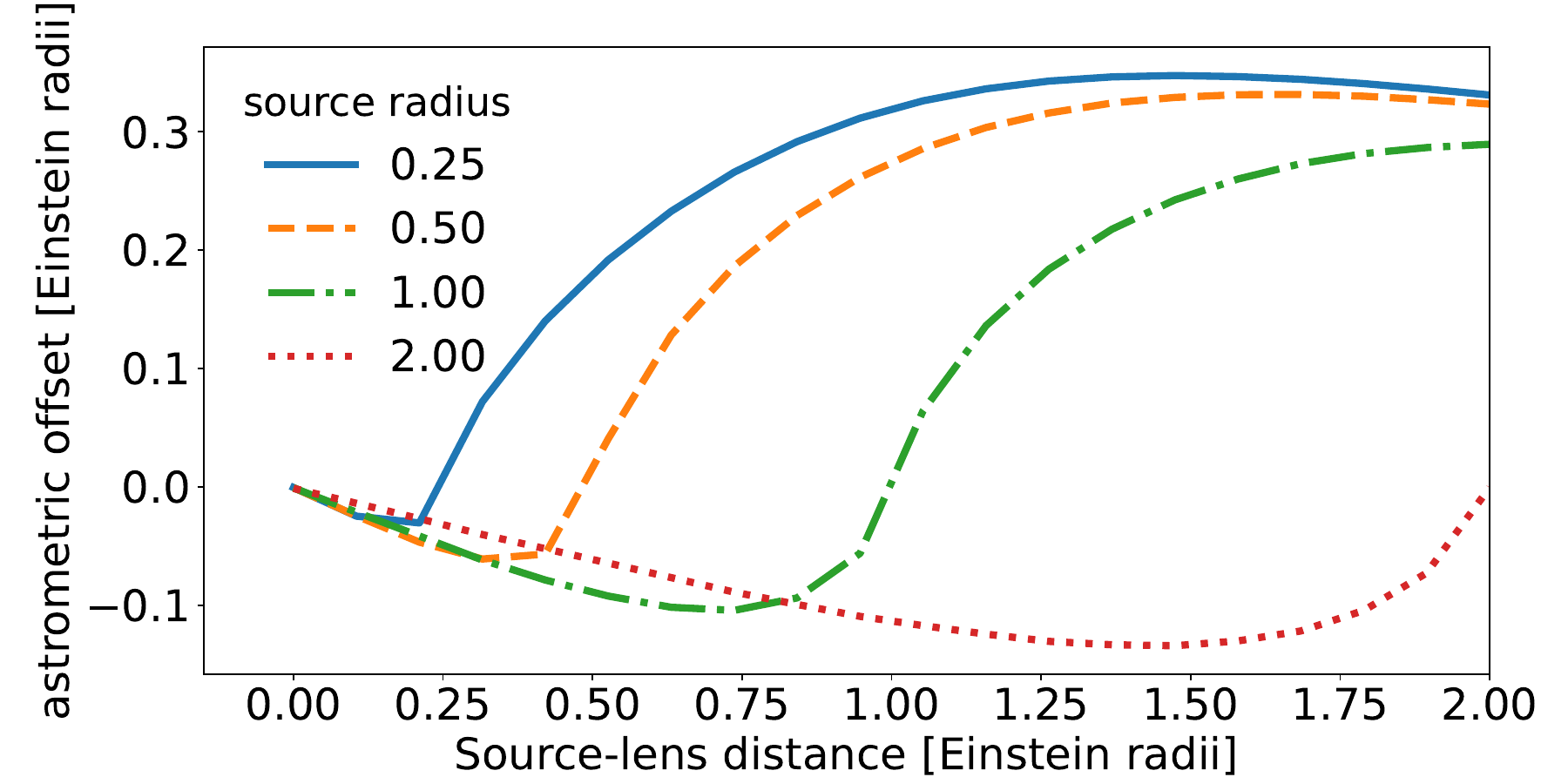}
    \caption{Astrometric offset as a function of distance between lens and source centers (on the image plane). 
             Several sizes of the source are shown, in units of Einstein radius. 
             The offsets reach a fraction of an Einstein radius, 
             which, for typical values of self-lensing systems, 
             would be about $10^{-4}$\,AU, 
             which is under a micro-arcsecond for systems
             at a distance of 100\,pc.
    }
    \label{fig: example astrometric shifts}
\end{figure}

In Figure~\ref{fig: example astrometric shifts} we show the astrometric offset
of sources of different sizes at various lens-source distances. 
The shifts are generally limited to a fraction of an Einstein radius. 
It should be noted that, for self-lensing systems, 
with lens mass of $\sim M_\odot$ and semimajor axis of $\sim 1$\,AU, 
the Einstein radius is about $10^{-4}$\,AU. 
At a distance of 100\,pc an offset of one Einstein radius 
would be equivalent to a micro-arcsecond of astrometric offset. 
Thus we conclude that self-lensing systems would have 
undetectable astrometric offsets. 

This method of calculation is quite computationally slow.\footnote{
On an Intel i7-9850H, calculating a lightcurve 
for a single system, including multiple geometries 
where the lens moves across the source,
can take several seconds. 
} 
To facilitate faster analysis we pre-calculate the relevant values
over a grid of parameters, and use them to interpolate the results
in between the grid points. 
In this work we utilized three pre-computed datasets, each covering a 
different range of parameters.
These parameters are summarized in Table~\ref{tab: matrix parameters}. 

\begin{deluxetable}{cccc}

    \centering
    \tabletypesize{\small}
    \tablewidth{0pt}
    \tablecolumns{4}
    \tablecaption{Pre-calculated datasets}
    \tablehead{\colhead{Source radii} & 
    \colhead{Distances} & 
    \colhead{Lens radii} &
    \colhead{Step sizes}}
    \startdata
    0.01 -- 0.1 & 0.0 -- 3.0  & 0.0 -- 2.0  & 0.005 \\
     0.1 -- 1.0 & 0.0 -- 10.0 & 0.0 -- 5.0  & 0.025 \\
     1.0 -- 5.0 & 0.0 -- 20.0 & 0.0 -- 10.0 & 0.1
    \enddata
    \tablecomments{All sizes are relative to the Einstein radius $R_E$. 
                   Step sizes for these datasets are uniform across the grid
                   in all three parameter axes. }
    \label{tab: matrix parameters}

\end{deluxetable}

For sources smaller than those available in the first dataset, 
i.e., $r_S<0.01$, 
we use the point source approximation given by Equation~\ref{eq: point source approximation}. 
For sources further away than the maximal distance, 
we also use the point source approximation, 
with the assumption that at $u\gg r_S$, 
each point on the source has a similar distance to the lens, 
so that the total magnifications is given by the same approximation. 
At these distances, the lensing effect yields only small magnifications (less than 1\%), 
so the exact choice of approximation does not affect our calculations in a major way. 
For sources larger than those in the last dataset, 
we use the solution of \cite{self_lensing_large_sources_Agol_2003}, 
using elliptical integrals. 

For each dataset, we only save the difference 
in flux and un-normalized moment, 
between geometries with adjacent source sizes. 
These results represent the fluxes and moments 
that would be measured for an annulus with uniform surface brightness. 
The reason we keep the results in this way, 
rather than store the results for disks of uniform brightness, 
is to be able to reconstruct the magnifications and offsets 
for different limb-darkening profiles. 
With a different relative surface brightness of each radius on the source disk, 
we can sum the contributions to the flux and moment for each annulus, 
using appropriate weighting, 
and produce a result for an arbitrary darkening profile. 
To recover the results for uniform disks, 
we simply sum the results of all annuli with equal weights.

\section{Physical properties of self-lensing systems}\label{sec: physical properties}

The lensing flare would appear different for various types of binary components. 
The source must be a luminous source, 
so it could be a main sequence or giant branch star (MS or GB) or a white dwarf (WD). 
In this work we do not address the MS/GB options, 
as the stars are substantially larger than WDs, 
and the signal would be diluted by a large factor 
unless considering very long orbits. 
Even for MS sources that have detectable flares,
e.g., for small stars with wide orbit companions 
(yielding large Einstein radii),
the signal would be swamped by false alarms from stellar flares. 
To simplify the search we focus on WD sources. 

The lens must be compact in order to generate a positive signal, 
so white dwarfs, neutron stars or black holes are the possible lenses.
The WD case is particularly challenging since the size of the physical object
is often comparable to the Einstein radius, 
so the signal is composed of a lensing flare 
superimposed on an eclipse dip. 
For NS or BH lenses, 
we assume the physical size of the lens is negligible 
and calculate only a positive flare. 
We simulate lenses of three types, with non-overlapping mass ranges:
WDs have $0.2 M_\odot <M_\text{lens}<1.2 M_\odot$, 
NSs have $1.2 M_\odot < M_\text{lens} < 2.5 M_\odot$
and BHs have $2.5 M_\odot < M_\text{lens} < 30 M_\odot$. 
It should be noted that in all our simulations and results, 
there is no difference between a NS and a BH, 
and they are treated as identical objects with different masses. 

We assume the radius of a WD is given by Equation 5 of \cite{self_lensing_Beskin_Tuntsov_2002} 
(see also \citealt{white_dwarf_mass_radius_relation_Nauenberg_1972}):
\begin{equation} \label{eq: white dwarf radius}
    R_\text{WD} = 0.01125 R_\odot \left[\left(\frac{M_\text{WD}}{1.454 M_\odot}\right)^{-\frac{2}{3}} - \left(\frac{M_\text{WD}}{1.454 M_\odot}\right)^{\frac{2}{3}}\right]^{\frac{1}{2}}.
\end{equation}

Another free parameter that determines the system properties
is the spatial distance between source and lens: the semimajor axis, $a$. 
The orbits we consider here are $10^{-2} \leq a \leq 10$\,AU 
for WD-WD binaries and $10^{-4} \leq a \leq 1$\,AU for the WD-NS and WD-BH binaries.
As a reference, two WDs with masses of 0.6\,$M_\odot$ at an orbit 
of $a=10^{-2}$ have a period of 8 hours. 
For $a=10^{-3}$ the period is 15 minutes, 
and for $a=10^{-4}$ the period is 30\,s. 
For one such WD orbiting a 30\,$M_\odot$ BH at a semimajor axis of $a=10^{-4}$
the orbital period is about 6\,s.
It should be noted that systems with $a=10^{-4}$\,AU 
will have a typical decay time due to gravitational radiation
\citep{white_dwarf_binaries_merge_Brown_2016}
on the order of a few years, or a few days, 
for a 0.6\,$M_\odot$ WD with a 1.5\,$M_\odot$ NS 
or a 30\,$M_\odot$ BH, respectively. 
Since such systems are expected to be very short lived, 
we do not expect to detect such systems. 

Although tighter orbits are physically allowed by the sizes of the objects, 
the smaller distances correspond to small Einstein radii, 
so the shortest period systems would be undetectable in most cases. 

The maximal semimajor-axes we consider in this work are $10$\,AU for WD-WD 
and only 1\,AU for the other binaries. 
Beyond 1\,AU the fraction of systems seen edge on, 
and the duty cycle of such systems, both decrease substantially. 
In addition, following up such systems with periods
on the order of a year becomes very difficult, 
so even if such a system is ever seen, 
it will most likely never be correctly identified. 

The temperatures of the WD sources, 
and lenses in the case of WD-WD systems, 
have a strong effect on the detectability of the system.
Hotter WDs will be brighter and can be detected 
at larger system-observer distance than cool ones. 
If the system is a double WD, 
the amount of dilution the flare of the lensed WD 
suffers from the light of the lensing (non-lensed) WD
is determined by the size and temperature of each component. 
The range of possible temperatures we use is $4000^\circ\text{K}\leq T_\text{WD}\leq 32000^\circ\text{K}$, 
which is similar to the range of values 
in the sample presented by \cite{white_dwarfs_in_Gaia_DR3_Fusillo_2021}. 
We assume the number of observable WDs with temperatures
above 32000$^\circ$K is small, and does not change our results substantially. 

We assume all systems follow circular orbits. 
This may be a good approximation for close binaries 
that have undergone common envelope evolution, 
and even in cases where the eccentricity is non-negligible, 
it has only a mild effect on the resulting flare. 
The exact distance and velocity of the lens at the time 
when it crosses near the line of sight of the source
would be different by some small factor from the values
we used here, with the assumption of a circular orbit. 
This can make the flare duration and size 
somewhat different from our simulations. 
But, since we are mostly interested in the average properties, 
we choose not to add the additional parameters
required to simulate eccentric orbits, 
which saves a large computational cost. 

In addition to the self-lensing optical signal, 
we also calculate the gravitational wave (GW) strain 
expected to be emitted from each binary in the simulation. 
The strain is given by: 
\begin{equation}\label{eq: gravitational strain}
    h_\text{GW} = 4 \frac{(G \mathcal{M})^{5/3}(\pi f)^{2/3}}{D_L c^4}, 
\end{equation}
where $\mathcal{M}=(M_sM_l)^{3/5}/(M_s+M_l)^{1/5}$ is the chirp mass, 
$f=2/P$ is the GW frequency calculated from the binary orbital period $P$, 
and $D_L$ is the luminosity distance (in case of Galactic binaries, this 
is simply the distance to the system). 

\begin{figure}
    \centering
    \includegraphics[width=1\linewidth]{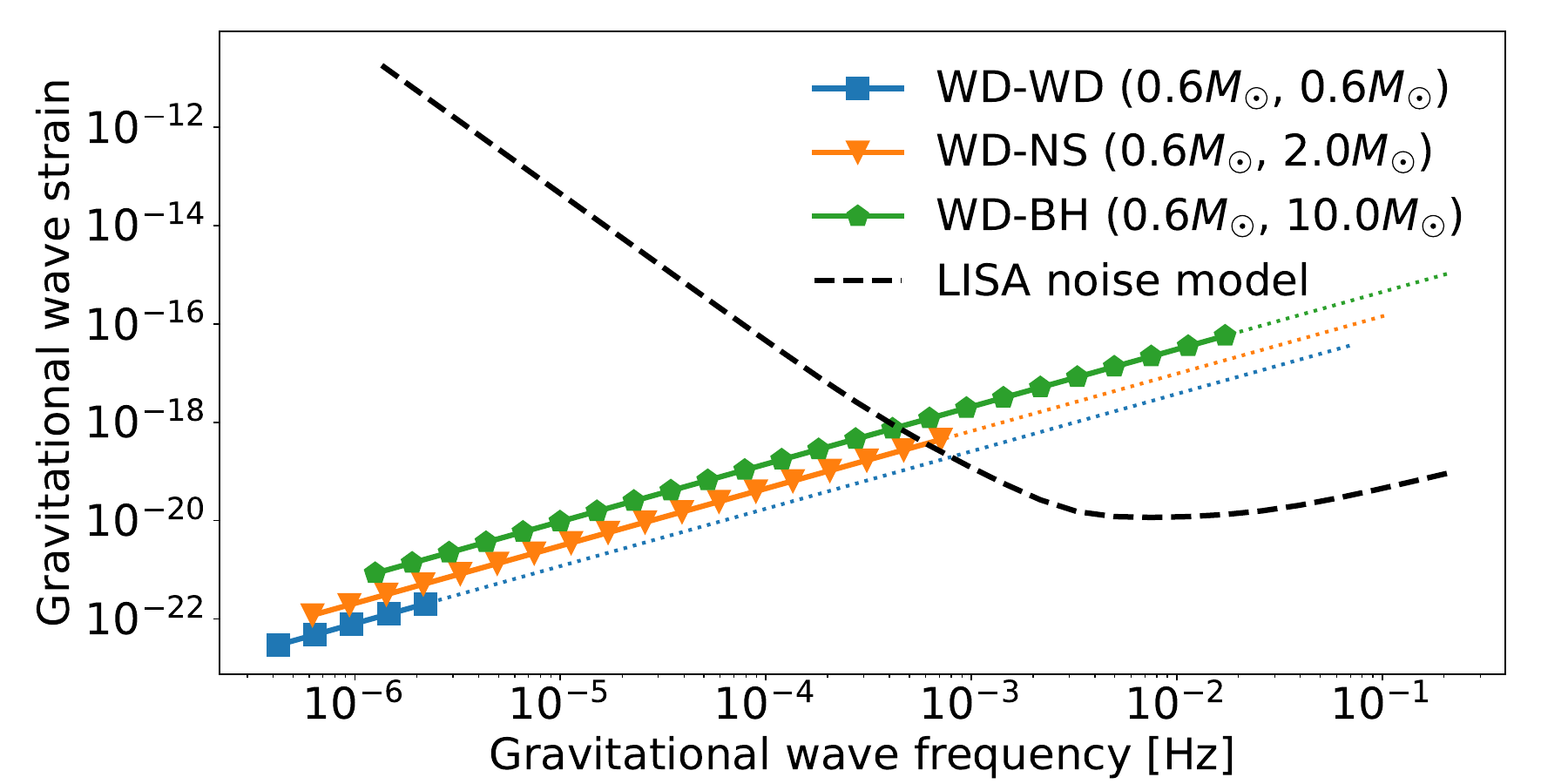}
    \caption{
        The expected gravitational strain, compared to the nominal LISA noise model,
        (given by \citealt{gravitational_wave_detectors_LISA_noise_model_Smith_2019}). 
        The DWD systems have equal mass of $M_l=M_s=0.6$\,M$_\odot$. 
        The WD-NS system has masses of $M_s=0.6$ and $M_l=2.0$\,M$_\odot$. 
        The WD-BH system has masses of $M_s=0.6$ and $M_l=10.0$\,M$_\odot$. 
        All systems are placed at 1\,kpc from the observer, 
        and we assume they are observed over 4 years of the LISA life time. 
        The solid lines represent systems have peak self-lensing flare amplitudes above 10\%. 
        The dotted lines show additional systems with lower orbits and stronger GW emission, 
        that have self-lensing flares below the 10\% level, and are harder to detect photometrically. 
        The dashed line is the LISA noise model. 
        We see that only the WD-BH systems can be detected easily by both 
        photometric all-sky surveys, and space-based GW detectors. 
    }
    \label{fig: gravitational strain}

\end{figure}

For self-lensing DWD binaries, this strain is well below the LISA 
noise level \citep{gravitational_wave_detectors_LISA_noise_model_Smith_2019}. 
This is in contrast to eclipsing WD-WD binaries that have much 
closer orbits and are excellent targets for LISA 
(e.g., \citealt{eclipsing_wds_shortest_orbit_Burdge_2019, eclipsing_wds_second_ZTF_Coughlin_2020, eclipsing_wds_second_shortest_Burdge_2020}). 
Short-period binaries with a NS or BH would be able to generate a detectable signal. 
Such measurements would provide additional constraints on the parameters of the system, 
or, conversely, could be used to narrow down the search for close binaries, 
if such systems are first detected by LISA at a distance where wide-field surveys 
cannot blindly detect the photometric variability. 
A plot of the expected LISA sensitivity, compared to a range of DWD, WD-NS and WD-BH
systems and the expected strain they would produce at a distance of 1\,kpc 
is shown in Figure~\ref{fig: gravitational strain}. 
We see that only massive WD-BH binaries have a chance of being detected
in both photometric surveys and space-based GW detectors. 

\section{Self-lensing parameter space}\label{sec: parameter space}

\begin{figure}
    \centering
    \includegraphics[width=1\linewidth]{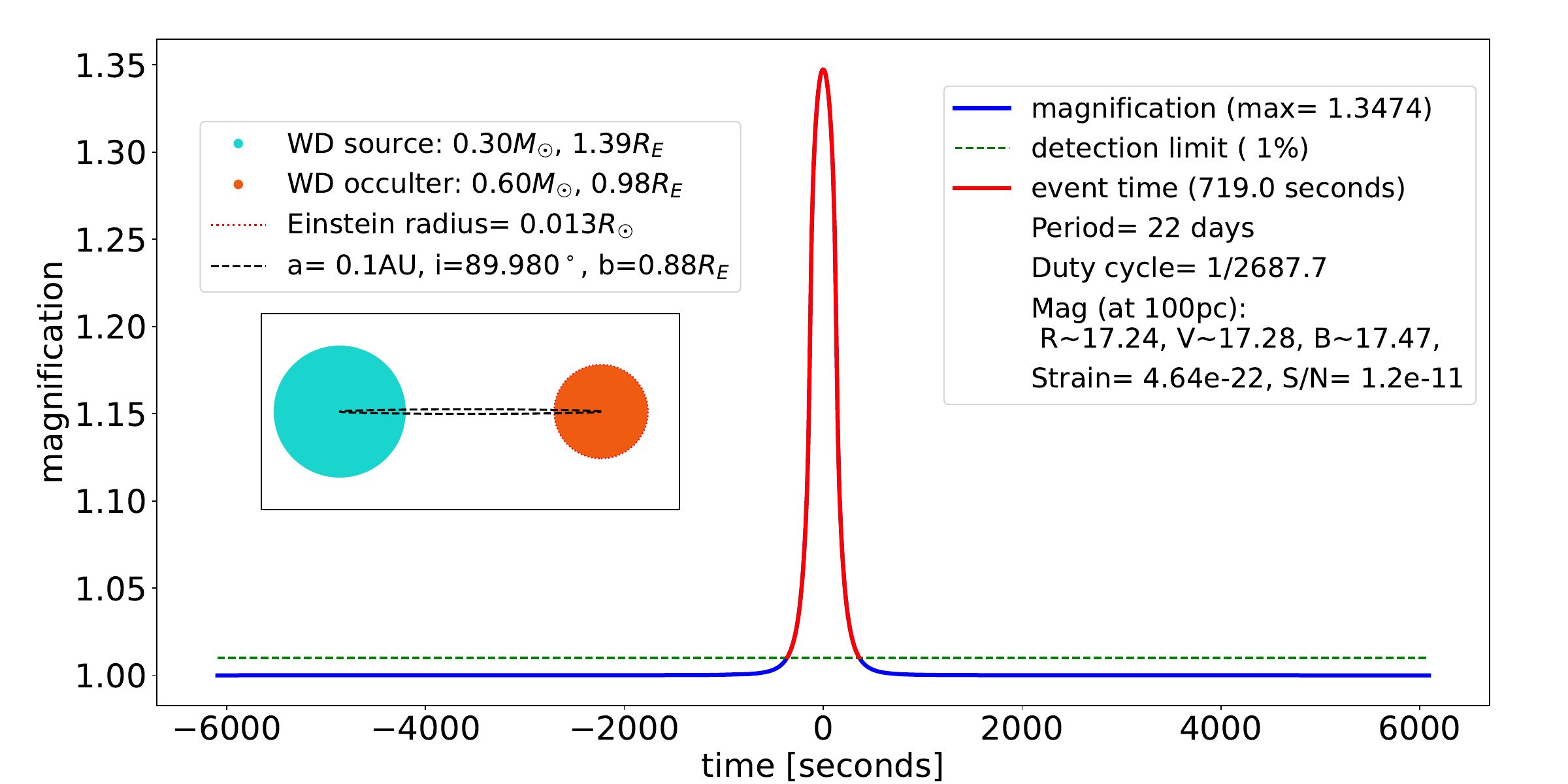}
    \caption{A simulated WD-WD system. 
             The main plot shows the magnification as a function of time, 
             while the inset shows a cartoon of the system
             (not to scale). 
             The masses of the stars are 0.3 and 0.6\,$M_\odot$. 
             The orbital radius is 0.1\,AU, 
             giving a 22 day orbit. 
             The larger and less massive WD is lensed
             by the smaller and more massive companion. 
             The flare duration is about 700\,s, 
             with a duty cycle of about $1/2700$. 
             The maximum magnification of 1.35, 
             along with the flare duration, 
             makes this an easy target to detect. 
             The low duty cycle and long period,
             however, will make it very hard to identify 
             such a system. 
             The gravitational strain of this system, of $h\approx 10^{-23}$ 
             is too weak to be detectable by, e.g., LISA. 
    }
    \label{fig: example WD-WD system}
\end{figure}

\begin{figure}
    \centering
    \includegraphics[width=1\linewidth]{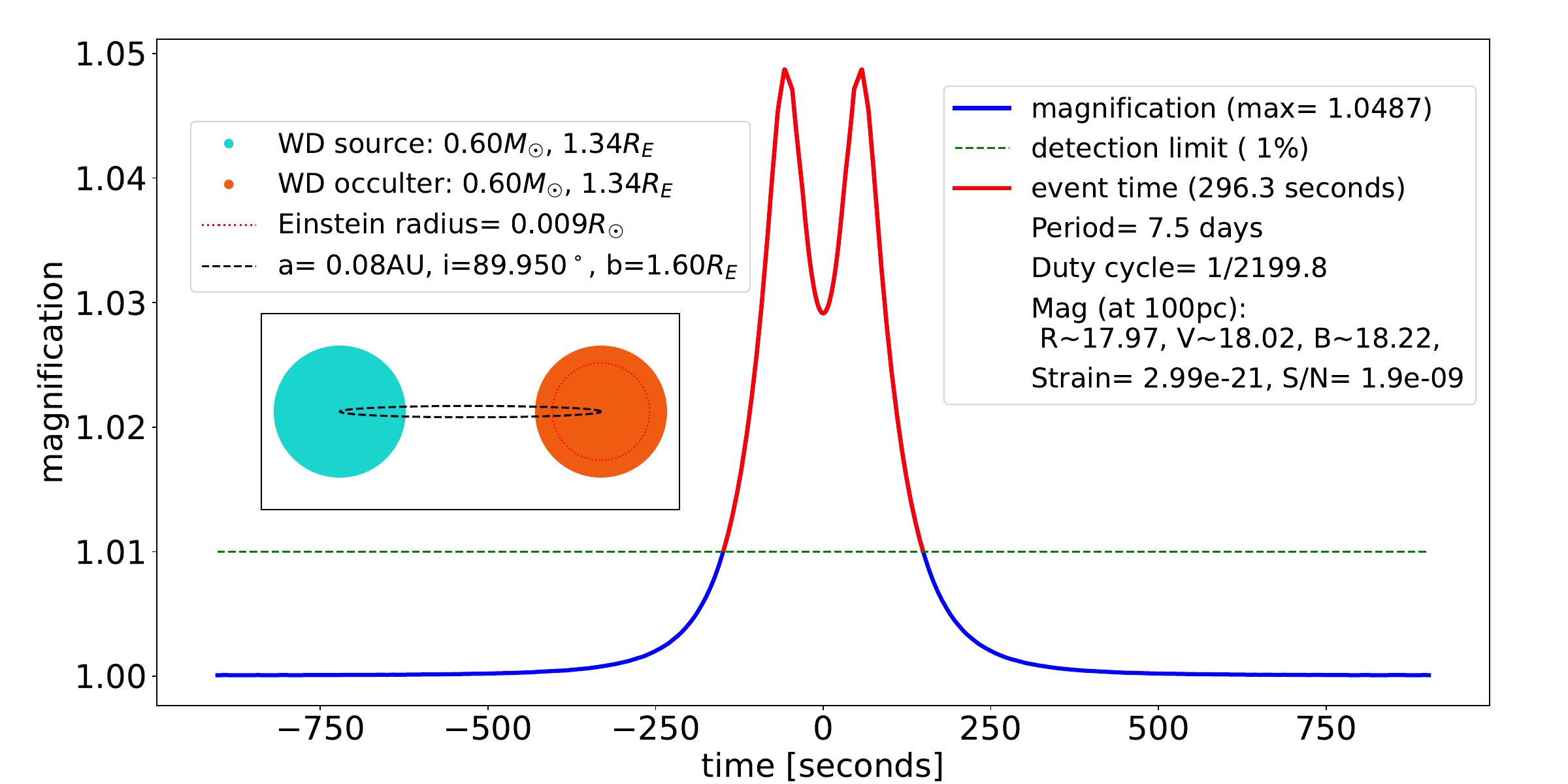}
    \caption{A simulated WD-WD system with a visible eclipse. 
             The plot is similar to Figure~\ref{fig: example WD-WD system}
             but with conditions that favor an eclipse in the middle 
             of the lensing flare. 
             The masses of the stars are both 0.6\,$M_\odot$, 
             such that the source is smaller than the previous example 
             shown in Figure~\ref{fig: example WD-WD system}
             The orbital radius is 0.08\,AU, 
             which also reduces the Einstein radius 
             relative to the previous example. 
             The increased source mass and closer orbit
             give a period of only 7.5 days. 
             The physical size of the occulting WD
             is seen to reduce the magnification during the 
             middle part of the flare. 
             The flare duration is about 300\,s, 
             with a duty cycle of about $1/2200$. 
             Since this system is symmetric, the 
             period would be halved and the duty cycle doubled, 
             compared to the parameters shown here. 
             The maximum magnification of only 1.05, 
             already makes this a challenging target
             for surveys not suited for high precision photometry. 
             The gravitational strain of this system, of $h\approx 10^{-22}$ 
             is too weak to be detectable by, e.g., LISA. 
    }
    \label{fig: example WD-WD system eclipse}
\end{figure}

\begin{figure}
    \centering
    \includegraphics[width=1\linewidth]{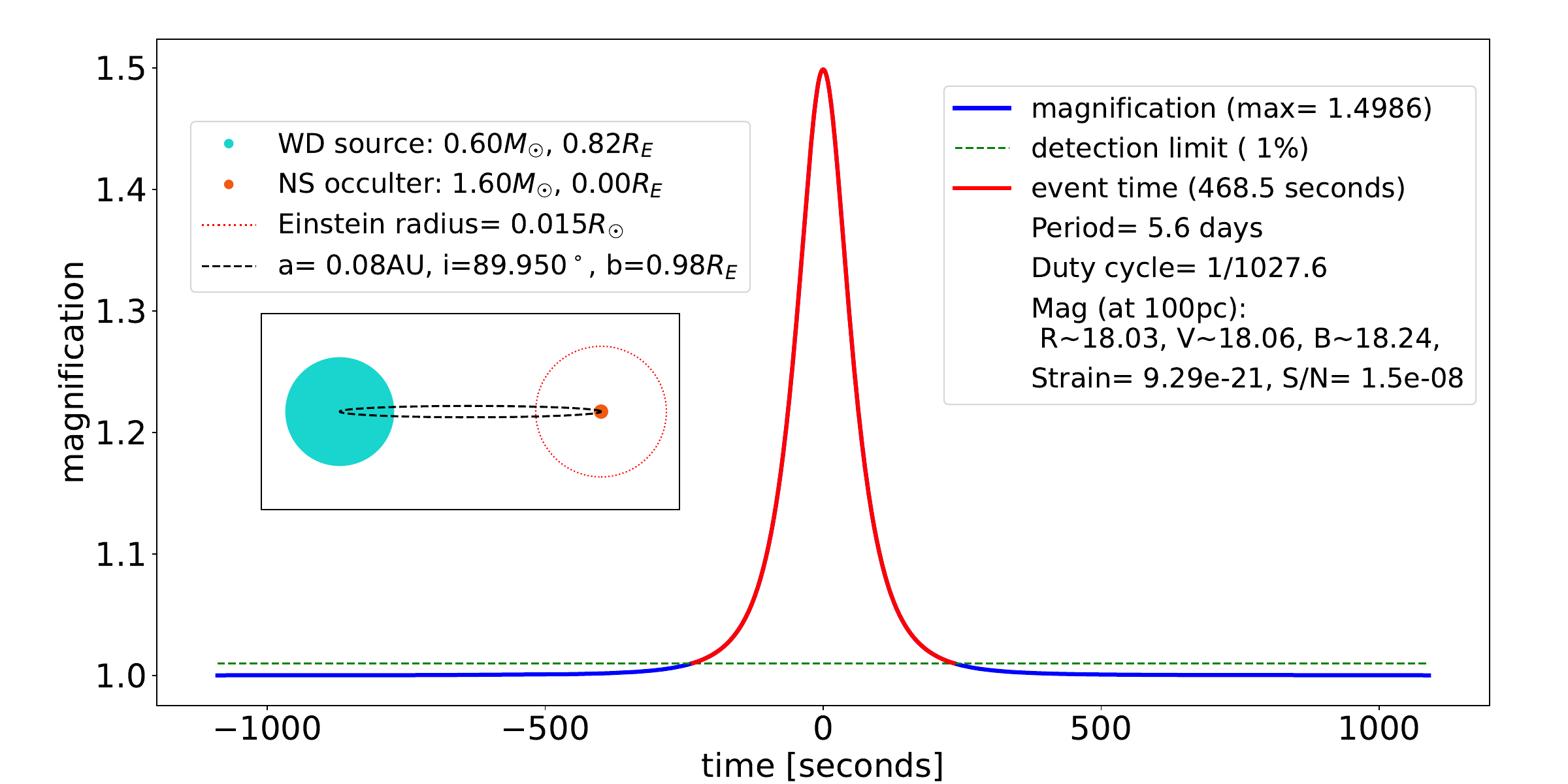}
    \caption{A simulated WD-NS system. 
             The plot is similar to Figure~\ref{fig: example WD-WD system}
             but with a NS lensing a WD.
             The masses of the stars are 0.6 and 1.6\,$M_\odot$. 
             In this example, 
             there is no eclipse, as the size of the NS
             (on the order of a few km) is negligible 
             when compared to the size of the WD 
             (on the order of thousands of km). 
             The orbital radius is 0.08\,AU, 
             giving a 5.6 day orbit. 
             The flare duration is about 500\,s, 
             with a duty cycle of about $1/1000$. 
             The maximum magnification of 1.5, 
             along with the flare duration, 
             makes this an easy target to detect.
             Since NSs are more massive and do not create eclipses, 
             they can be detected at lower separations. 
             Thus, there may be detectable WD-NS systems with 
             a substantially higher duty cycle
             and shorter flare repeat period. 
             The much lower brightness of the NS also 
             means there is no visible lensing when the 
             WD moves in front of the NS. 
             The gravitational strain of this system, of $h\approx 10^{-22}$ 
             is too weak to be detectable by, e.g., LISA. 
    }
    \label{fig: example WD-NS system}
\end{figure}

\begin{figure}
    \centering
    \includegraphics[width=1\linewidth]{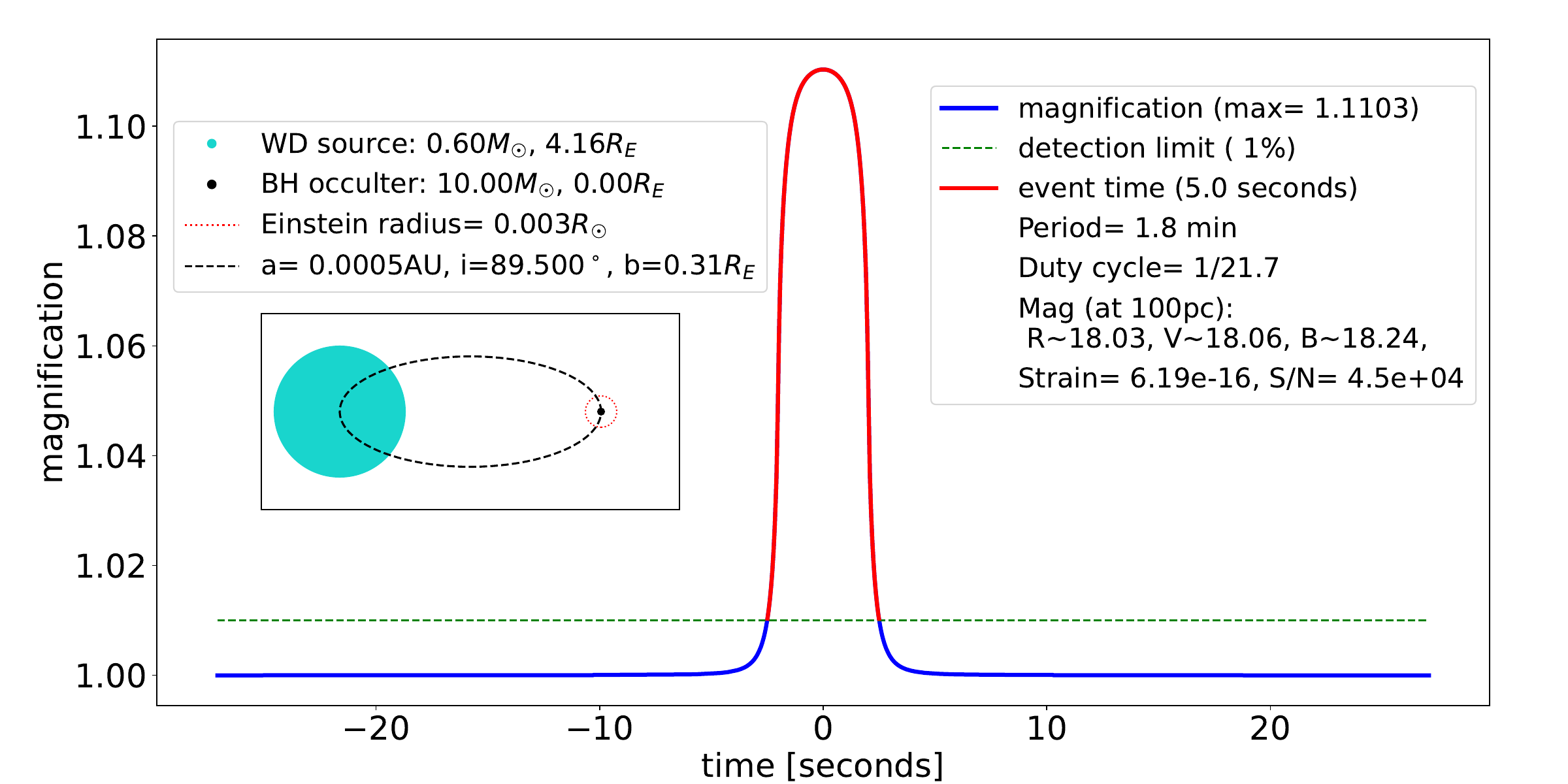}
    \caption{A simulated WD-BH system. 
             The plot is similar to Figure~\ref{fig: example WD-WD system}
             but with a BH lensing a WD.
             The masses of the stars are 0.6 and 10.0\,$M_\odot$. 
             In this case, similar to the previous example, there is no eclipse. 
             The high mass of the lens makes it possible 
             to detect flares at lower orbital radius and inclination values. 
             The orbit is only $5\times 10^{-4}$\,AU, 
             giving a 1.8 minute period. 
             The flare duration is about 5\,s, 
             with a duty cycle of about $1/22$.
             The maximum magnification of 1.1, 
             along with the short repeat time, 
             makes this a good target for 
             high cadence surveys. 
             While BHs are expected to be rare, 
             the higher duty cycle, and larger
             range of inclination angles 
             that are still detectable, 
             might make these sorts of targets 
             favorable despite the presumed 
             lower abundance of systems. 
             The gravitational strain of this system, of $h\approx 10^{-19}$ 
             is easily detected by a gravitational wave detector like LISA. 
    }
    \label{fig: example WD-BH system}
\end{figure}

Some example simulated systems are shown in 
Figures~\ref{fig: example WD-WD system}--\ref{fig: example WD-BH system}. 
We show the resulting lightcurves 
in a time range close to the time of the flare
(not the entire period, which is usually
much longer than the flare duration). 
Note that at different masses and orbits
the typical time scales and magnifications can vary dramatically. 
The parameters for each systems are noted in two boxes overlaid
on the plot. These include, among other information, 
the inclination $i$, which is defined as edge on when $i=90$, 
and the impact parameter $b$, which is the on-sky distance between
lens and source centers at the point of closest approach.

\begin{figure*}
    \centering
    \includegraphics[width=1.0\linewidth]{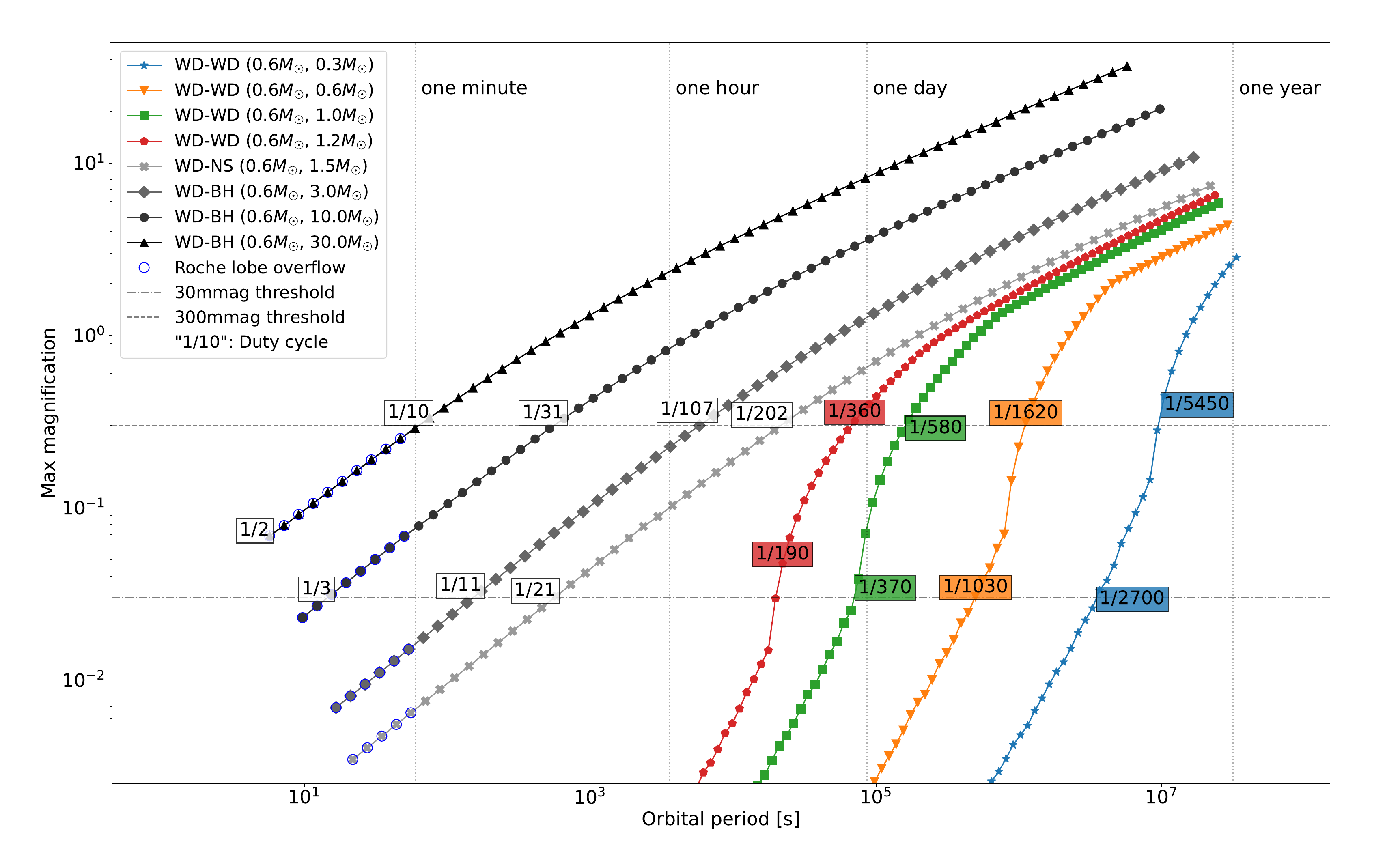}
    \caption{Parameter space of self-lensing flares. 
             The magnification vs.~orbital period for 
             self-lensing systems with different lens masses. 
             All systems have a source WD with a mass of 0.6\,$M_\odot$. 
             The colored curves are for WD-WD systems with masses 
             ranging 0.3 -- 1.2\,$M_\odot$, 
             while the grey and black curves represent 
             WD-NS and WD-BH systems with masses 
             in the range 1.5 -- 30\,$M_\odot$. 
             The two dashed, horizontal lines
             represent 30 and 300\,mmag detection thresholds. 
             The text labels show the duty cycle
             (and are a good approximation for the geometric factor)
             for each system at the period where it is above 
             each of the thresholds. 
             These numbers give a rough estimate of the 
             number of systems and the number of observations
             needed per system to see a flare. 
             The blue circles around short period systems
             represent systems undergoing Roche-lobe overflow. 
    }
    \label{fig: mag vs period}
\end{figure*}

As discussed above, the different parameters of 
each binary have different and interacting effects
on the detection prospects for various surveys. 
The biggest considerations are the lens mass
and the orbital radius. 
These determine the Einstein radius 
which sets the size of the flares.
If we assume constant masses for the objects, 
the semimajor axis, and so the orbital period, 
determines the flare size. 
On the other hand, shorter periods are
easier to detect, because of the higher duty cycle
and larger geometric angle in which lensing can be seen. 
In Figure~\ref{fig: mag vs period} we show the magnification
as a function of orbital period. 
The colored curves show WD-WD systems with 
lensing WDs of different masses. 
The grey and black curves show a WD with
a NS or a BH with lenses of different masses. 
In all cases the source is a typical WD 
of 0.6\,$M_\odot$. 

The WD-WD systems have a sharp drop-off at low periods, 
which correspond to where the lensing WD starts to also 
occult the source, which puts them effectively into 
the regime of eclipsing white dwarfs. 
This happens at orbital periods of a few hours to 
close to a year, depending on mass. 
It should be noted that the majority of eclipsing WDs
have been found with considerably shorter periods, 
owing to the same considerations of duty cycle and geometry
that affect self-lensing
\citep{eclipsing_wds_shortest_orbit_Burdge_2019, eclipsing_wds_second_ZTF_Coughlin_2020, eclipsing_wds_second_shortest_Burdge_2020}. 

The WD-NS and WD-BH systems do not show such a drop-off
as their physical size is negligible compared to 
the source size and the Einstein radius. 
Their flares are visible above even the
higher threshold of 300\,mmag down to 
a period of hours (for the 1.5\,$M_\odot$ NS) 
or even minutes (for the 30\,$M_\odot$ BH). 

The text labels on the curves specify the duty cycle
for each system as it crosses either the 
30\,mmag or 300\,mmag thresholds. 
Less massive lenses cross these thresholds
at relatively longer periods, 
corresponding to lower duty cycles. 
The geometric angle where the flare is visible
has a value similar to the duty cycle. 
Thus an equal mass WD-WD system with 0.6\,$M_\odot$ 
(the orange triangles in the figure)
can be detected with 300\,mmag precision at a period of 12.7 days, 
with a duty cycle of 1/1620.
Thus only one in $\sim 1600$ systems will have
the correct inclination to be visible, 
and only one in $\sim 1600$ observations 
will overlap with the flare time
(assuming the exposure time is shorter than the flare duration, 
which is $\approx 700$\,s in this case).
A system with a more massive WD lens 
with a mass of 1.2\,$M_\odot$ 
(the red pentagons in the figure)
crosses the 300\,mmag threshold at about 1 day, 
with a duty cycle (and geometric factor) of 1/360. 
This is still a rare occurrence, 
and is complicated by the relative scarcity 
of such heavy WDs. 
If a population of massive WDs exists
but remains undetected due to the objects' smaller size, 
self lensing would be an excellent way to detect it, 
if a sufficient number of binary WDs can be observed
with sufficient cadence
to overcome the 1/360 duty cycle. 

For heavier lenses, 
that also do not eclipse the source, 
the part of the parameter space that
is most interesting for detection is 
that of minutes to hours periods. 
Such systems have duty cycles between
1/10 and 1/100 at the periods where they 
are still above threshold, 
and any candidate systems found 
could be followed up easily with a few hours
of high cadence imaging. 
The limiting factor on the 
intrinsic numbers of these systems 
comes from the size of the orbits: 
at orbits shorter than about one minute
the WD undergoes Roche-lobe overflow and 
starts to spill mass unto the lens. 
For orbits of a few minutes, 
before mass transfer begins, 
gravitational wave emission will 
shrink the orbits and deplete the number 
of systems available at any given time. 
Thus, the most viable region of the parameter space
to observe NS and BH lenses around a WD, 
is that of 10--100 minute orbits. 
The flare durations in such cases 
are on the order of 10--100 seconds.

\section{Detection prospects and survey strategies}\label{sec: survey strategies}

Different surveys have different observing strategies, 
that can dramatically change the detection prospects 
for self-lensing events. 
The most significant parameters for determining
the detection rate would be the field of view,
the depth, the number of visits, 
and the photometric precision as a function of magnitude. 
The field of view determines the number of stars
that are observed in each visit. 
Surveys sensitive to fainter objects will also 
monitor many more stars at higher distances, 
compared to shallower surveys. 
For each set of parameters, 
the probability that the survey observes
the system during a flare is proportional 
to the total amount of time the survey 
observes each field 
(i.e., the exposure time multiplied by the number of visits). 
The third factor, 
the photometric precision, 
determines the fainest flares that can be identified by a survey. 
Fainter flares correspond to systems with closer orbits, 
that have a higher probability to be seen with the 
right timing and inclination. 

We have chosen to include several surveys that have relatively 
high cadence, and either a large sky coverage or large depth. 
This list is not exhaustive but we believe it is representative
of some common strategies for existing and planned surveys. 

ZTF is a large area, medium depth, all (Northern) sky survey
that is focused on finding transients with time scales of days 
\citep{Zwicky_transient_facility_Bellm_Kulkarni_2019}.
TESS is a satellite mission designed for detecting transits of planets
around other stars. It covers the entire sky with long series of observations
of 120\,s, but has small apertures on its array of four cameras, 
with a low limiting magnitude and is thus limited to nearby targets 
\citep{transiting_exoplanet_survey_satellite_Ricker_2014}. 
LSST is a planned large scale survey of the Southern sky using 
an unprecedented combination of a large aperture and large field of view, 
which gives it an advantage in terms of space volume covered 
\citep{large_synoptic_survey_telescope_Ivezic_2019}. 
The Dark Energy Camera (DECam; \citealt{decam_dark_energy_camera_DePy_2008}) has a large aperture but 
a smaller field of view relative to other surveys we consider. 
The DECam Deep Drilling Fields (DDF) survey is a 3-night cadence survey
with multiple exposures in a series, that may be useful in identifying 
self-lensing flares \citep{decam_deep_drilling_field_DDF_Graham_2023}. 
CuRIOS is a planned array of cubesats that would observe a large fraction
of the sky continuously on minutes time-scales, to a depth 
substantially larger than that of TESS, 
and could prove effective in detecting and identifying self-lensing flares
\citep{curios_slides_spie_Lu_Gulick_2022}. 
Finally, LAST is a ground array of small aperture telescopes
that also combines a large field of view with medium depth
and repeated exposures on time scales of minutes 
\citep{LAST_telescope_array_system_overview_Ofek_2023}. 
The key properties of each survey as they relate to finding self-lensing flares
are shown in Table~\ref{tab: survey parameters}. 
Note that for surveys with multiple filters and exposure times
we have chosen a representative middle value to use in all simulations
(which simplified the process). 
As we are interested in mean values and rough estimates
for the number of detections, the exact response of each 
filter to each system color is not critical, only the average values. 
In the table, we spell out the different filters for each survey but 
highlight the one used in the simulation in bold.

\begin{deluxetable*}{ccccccccc}

    \centering
    \tabletypesize{\small}
    \tablewidth{0pt}
    \tablecolumns{9}
    \tablecaption{Survey parameters}
    \tablehead{
    \colhead{Name} & 
    \colhead{Exp.~time} &
    \colhead{Series length} &
    \colhead{Field of view} &
    \colhead{Aperture} &
    \colhead{Lim.~mag.} &
    \colhead{Filters} &
    \colhead{phot.~prec.} &     
    \colhead{Duty cycle} \\
    \colhead{} &
    \colhead{(seconds)} &
    \colhead{} &
    \colhead{(deg$^2$)} &
    \colhead{(meters)} &
    \colhead{} &
    \colhead{} &
    \colhead{(percent)} &
    \colhead{(percent)} 
    }
    \startdata 
    TESS                  & 120  & $\sim 10^5$ & $4\times 24$       & 0.1  & 16   & broadband                  & 0.1--10\% & $\sim 100$\%  \\
    ZTF                   & 30   & 1           & 37                 & 1.2  & 21   & g, \textbf{r}, i           & 1--1\%    & $\sim 25$\%   \\
    LSST                  & 15   & 2           & 10                 & 8.0  & 24   & u, \textbf{g}, r, i, z, y  & 1--10\%   & $\sim 25$\%   \\
    DECam DDF             & 100  & 15          & 3                  & 4.0  & 23   & g, \textbf{r}, i           & 1--10\%   & $\sim 25$\%   \\
    CuRIOS                & 30   & 30          & 10                 & 0.1  & 20   & r                          & 0.5--20\% & $\sim 100$\%  \\
    LAST                  & 15   & 20          & $36 \times 7.4$    & 0.28 & 20   & broadband                  & 2--15\%   & $\sim 25$\%    
    \enddata
    \tablecomments{Different surveys used in the simulations. 
    To simplify calculations, and get a rough estimate for the number of detections in each survey,
    we used only a single filter in the simulations. We adopted the cadence of all filters combined, 
    under the assumption that a detection in any filter can be later followed up as a candidate self-lensing flares. 
    E.g., if ZTF takes two images in $r$ and $g$ in one night, our simulations will assume two images were taken in $r$, 
    when calculating the cadence and detection probabilities. 
    The filter used in the simulation is highlighted in bold in the table. 
    The duty cycle for ground based observatories accounts for loss of observations due to daylight, 
    bad weather, and assumes an average of 6h per night of useful observations. }
    \label{tab: survey parameters}

\end{deluxetable*}

\begin{figure}
    \centering
    \includegraphics[width=1\linewidth]{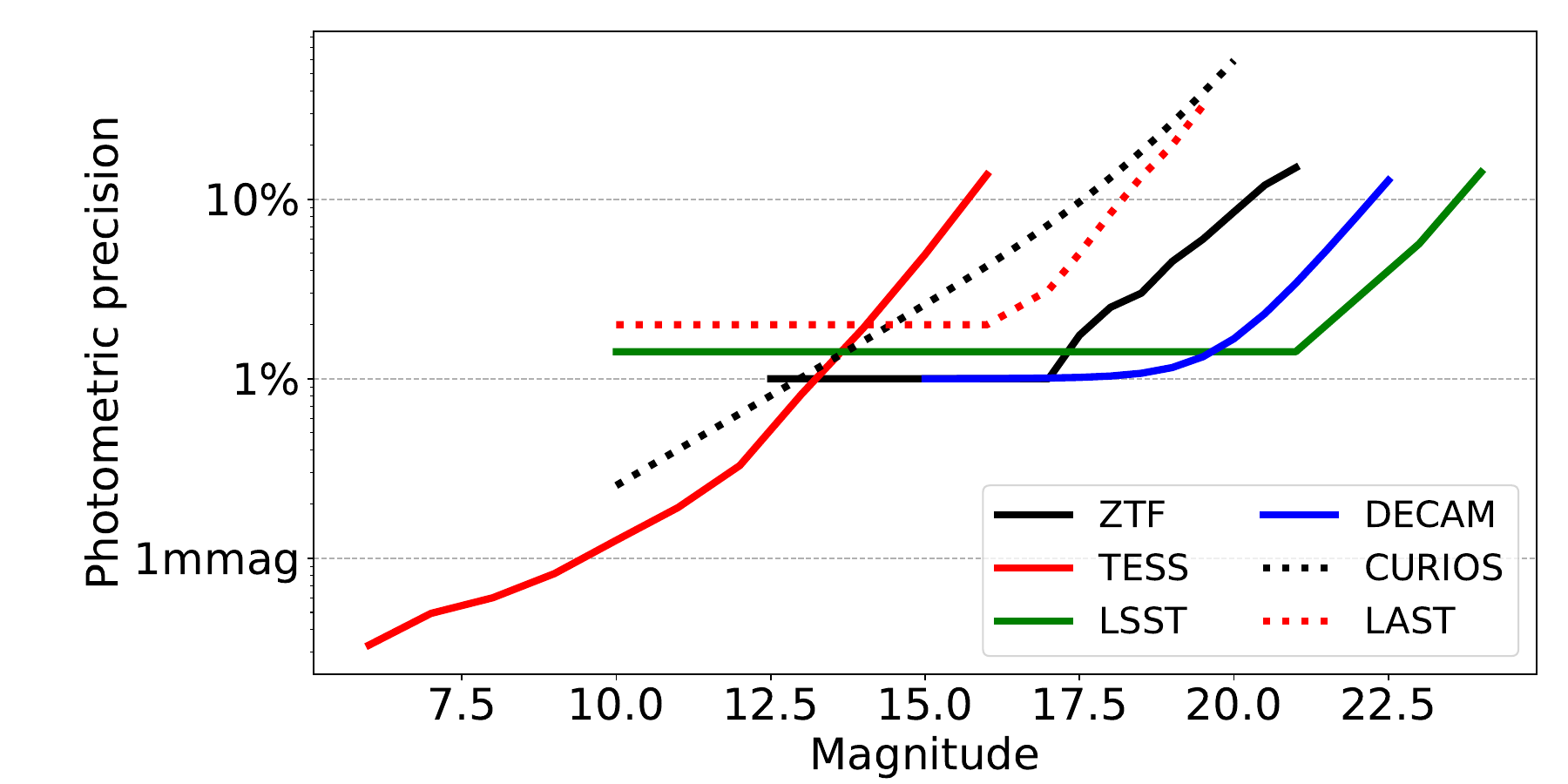}
    \caption{Photometric precision of various surveys used in the simulations, 
             as a function of the magnitude of observed objects. }
    \label{fig: photometric precision}
\end{figure}

To decide if a survey can detect a certain flare at a given distance
we use estimates for the precision of each survey as a function of apparent magnitude. 
The photometric precision used in the simulations
for various surveys are shown in Figure~\ref{fig: photometric precision}. 
The precision values for ZTF are taken from \cite{zwicky_transient_facility_data_processing_Masci_2019}, Figure 9. 
The precision values for TESS are from a technical note.\footnote{
\url{https://heasarc.gsfc.nasa.gov/docs/tess/observing-technical.html}. 
The precision is adapted by estimating the middle of the distribution 
of red points, scaled down from 1-hour exposures by $\sqrt{60/2}$ 
as an estimate for the precision at 2 minute exposure time. 
}
The precision values for LSST are given in \cite{large_synoptic_survey_telescope_Ivezic_2019}, Table 3, 
multiplying the value in the $\sigma_1$ column by $\sqrt{2}$
to convert from two to one exposure of 15\,s. 
We also assume that for stars brighter than 20th magnitude, 
the single exposure precision never increases above 1.4\%. 
The precision values for DECam are estimated by a noise model 
with a background count of 20\,e$^{-}$\,s$^{-1}$, 
a read noise of 12\,e$^{-}$ per exposure, 
an aperture of 4\,m, 
exposure times of 100\,s and a maximum precision of 1\%. 
The precision values for CuRIOS are given by assuming
a 15\,cm aperture and read-noise of $\sim 1$\,e$^{-}$ per pixel per frame, 
based on a preliminary design using the Sony IMS455 sensor (private communication). 
The precision values for LAST are given by \cite{LAST_telescope_array_system_overview_Ofek_2023}, Figure 16, 
blue curve. 
For stars brighter than 16.5 we assumed the precision does not 
improve beyond 2\% (S/N$\approx 50$). 

For surveys that take multiple exposures in a single visit 
(a photometric series), 
and for flares that are not shorter than a single exposure, 
the chances of detecting a flare in each visit would increase
compared to the single epoch detection, 
as S/N can be accumulated from multiple images.
If the series is longer than the entire period of the system, 
there could be multiple flares detected consecutively. 
Such detections would make it much easier to identify
the flare is periodic, but we do not address this 
in the current set of simulations, 
only the chance to detect a flare in general. 

These considerations illustrate that it is not trivial 
to estimate the detection rate, 
unless considering all the properties of the survey and the 
individual lensing system. 
Thus, we calculate the mean probability 
for detecting each system, 
under the observing parameters of each survey, 
for various distances (in logarithmic steps). 
The distance to the system determines its 
apparent magnitude which sets the photometric precision. 
We test further distances until the system moves 
below the detection threshold (fainter than the limiting magnitude). 

For each system, survey and distance, 
the detection probability for a single visit $P_\text{visit}$ is calculated. 
We calculate from that the total probability to detect a flare in a single field $P_\text{field}$
after a number of visits, $N_\text{visits}$, 
which is the average number of times
each field is observed over the life of the survey. 
The probability per field is then given by
\begin{equation}
    P_\text{field} = 1 - (1 - P_\text{visit})^{N_\text{visits}}. 
\end{equation}

This probability, which is a function of distance, 
is multiplied by the volume spanned by the field of view
area of all the fields in the survey, 
times the difference in distances. 
This defines an \emph{effective volume} for the survey:
\begin{equation}
    V_\text{effective} = N_\text{fields} \sum_{r=r_\text{min}}^{r_\text{max}} P_\text{field} \Delta r, 
\end{equation}
where the distance $r$ takes values starting from 3\,pc up to the maximum distance
that allows the system to be observed. 
Multiplying the effective volume 
by the expected space density of such systems,
including random inclinations, 
gives an estimate of the event rate. 

It is also possible to find the single-visit detection probability 
for specific object parameters, 
and compare that with systems that have already been observed, 
where the distance and some physical properties are known. 
This can be used to place limits on 
the presence of unseen companions. 

The single visit detection probability
is a key metric that must be estimated 
for each system, survey, and distance. 
Once the system parameters are determined, 
we can calculate the flare shape and intensity, 
and apply each survey to the same flare in turn. 
For each survey, the combination of precision, 
exposure time, number of images per visits (series length) 
and detector dead time is different, 
and yields a different signal 
when applied to each flare. 

If the flare duration is much shorter than the exposure time 
($t_\text{flare} < 0.1 t_\text{exp}$), 
the signal is given by the average flare intensity, 
over the exposure time. 
We calculate the signal-to-noise ratio (S/N) 
by dividing the signal with the noise level $\sigma_p$ 
(equivalent to the photometric precision) at each distance. 
\begin{equation}\label{eq: snr diluted}
    \text{S/N} = \frac{\int f(t) dt}{t_\text{exp} \sigma_p},
\end{equation}
where $f(t) \equiv M(t) - 1$ is the flare flux above the 
constant star light, given by the magnification $M(t)$ minus one. 
Both $f(t)$ and $\sigma_p$ are relative to the constant star's flux.

If the exposure time is not much longer than the flare, 
we apply a sliding window approach to finding the S/N. 
We build a uniform window spanning the exposure time, 
and convolve that with the lightcurve:
\begin{equation}\label{eq: snr sliding window}
    \text{S/N}(t) = \frac{1}{\sigma_p}(H \star f)(t)  = \frac{1}{\sigma_p} \int H(t-t')f(t') dt',
\end{equation}
where $\square \star \square$ is the convolution operator, 
and $H(t)$ is the ``hat function" defined by

\begin{equation}\label{eq: hat function}
    H(t) = \begin{cases} 1/t_\text{exp} & -\tfrac{1}{2} t_\text{exp}< t <\tfrac{1}{2} t_\text{exp} \\ 0 & \text{otherwise}.\end{cases}
\end{equation}
This takes into consideration all possible
time offsets between the exposure and the flare, 
and produces different values of S/N 
when integrating different parts of the flare. 

This result is for a single exposure. 
If the series is longer than that, 
we shift the resulting S/N$(t)$ function 
in time by intervals of $\Delta t = t_\text{exp}+t_\text{dead}$ 
and add the results from one exposure to the other
in quadrature, 
\begin{multline}\label{eq: snr sliding series}
    \text{S/N}_\text{total} = \left[\text{S/N}(t)^2 + \text{S/N}(t+\Delta t)^2 \right. \\ \left. +\text{S/N}(t+2\Delta t)^2 + \ldots \right]^{1/2}.
\end{multline}
This accounts for multiple exposures spanning a single flare, 
applying a matched-filter approach to adding the S/N from separate exposures. 

To convert the S/N into a detection probability, 
we assume the noise in each exposure is normally distributed,
and find the probability of each S/N value of being detected
above a pre-defined threshold of $\tau=5$:
\begin{equation}\label{eq: error function probability}
    P_\text{flare} = \frac{1}{2}\left[1 + Er(\text{S/N} - \tau) \right], 
\end{equation}
using the error function:
\begin{equation}\label{eq: error function}
    Er(z) = \frac{2}{\pi}\int_0^z e^{-x^2}dx.
\end{equation}
For S/N~$\ll\tau$ the probability is negligible, 
while for S/N~$\gg\tau$ the probability approaches 1, 
which happens for bright flares. 
For S/N~$\sim \tau$ we get the probability
to detect such a flare, with a S/N that fluctuates 
with a normal noise with standard deviation of one. 

When the S/N is time-dependent, 
e.g., Equation~\ref{eq: snr sliding window}, 
we calculate the probability for each time bin of S/N$(t)$, 
and average over the entire period to find the 
total probability assuming the phase of the orbit
is uniformly distributed relative to the time of the exposure/series. 

If a single visit with multiple exposures
is long enough to observe multiple flares, 
we calculate the probability to have $N_\text{flares}$ in the series. 
For each system this spans two numbers, offset by one flare. 
The probability to detect any flares is:
\begin{equation}
    P_\text{multi flare} = 1 - (1 - P_\text{flare})^{N_\text{flare}}, 
\end{equation}
and the total visit probability is the weighted sum of the 
results for the two possible number of flares.

\section{Results}\label{sec: results}

One way to summarize the results of each survey over many 
different self-lensing systems is to calculate the effective volume
given some set of system parameters. 
The effective volume is the product of the space volume observed
(the field of view, multiplied by the distance range 
where the systems were placed in the simulation) 
with the probability to detect a single flare from such a system 
over the life-time of the survey. 
Our simulations assume all surveys observe for 5 years, 
which is representative of the example surveys we used
and allows us to compare each survey's yield per unit time. 

The effective volume is necessarily a function of all the system parameters, 
but we are mostly interested in the marginalized effective volume, 
where we choose some reasonable values for the distribution of WDs, NSs and BHs, 
and calculate the mean effective volume for those systems. 

The inclination of different systems is uniform in $\cos i$, 
due to the geometry of binaries and is model-independent. 
The WD masses are chosen using a Gaussian distribution 
centered on $\mu_{M_{WD}}=0.6$, with a standard deviation of
$\sigma_{M_{WD}}=0.2$, both in units of Solar mass. 
For WD-WD models we use the same mass distribution for both source and lens. 
The temperatures of the WDs are chosen from a power-law distribution with 
an index of $q_{T_{WD}}=-2$. 
For NSs and BHs we assume a power law distribution of masses with an
index $q_{M_{BH}}=-2.3$. 
We will later show how the results change when modifying this value. 

\begin{figure}
    \centering
    \includegraphics[width=1\linewidth]{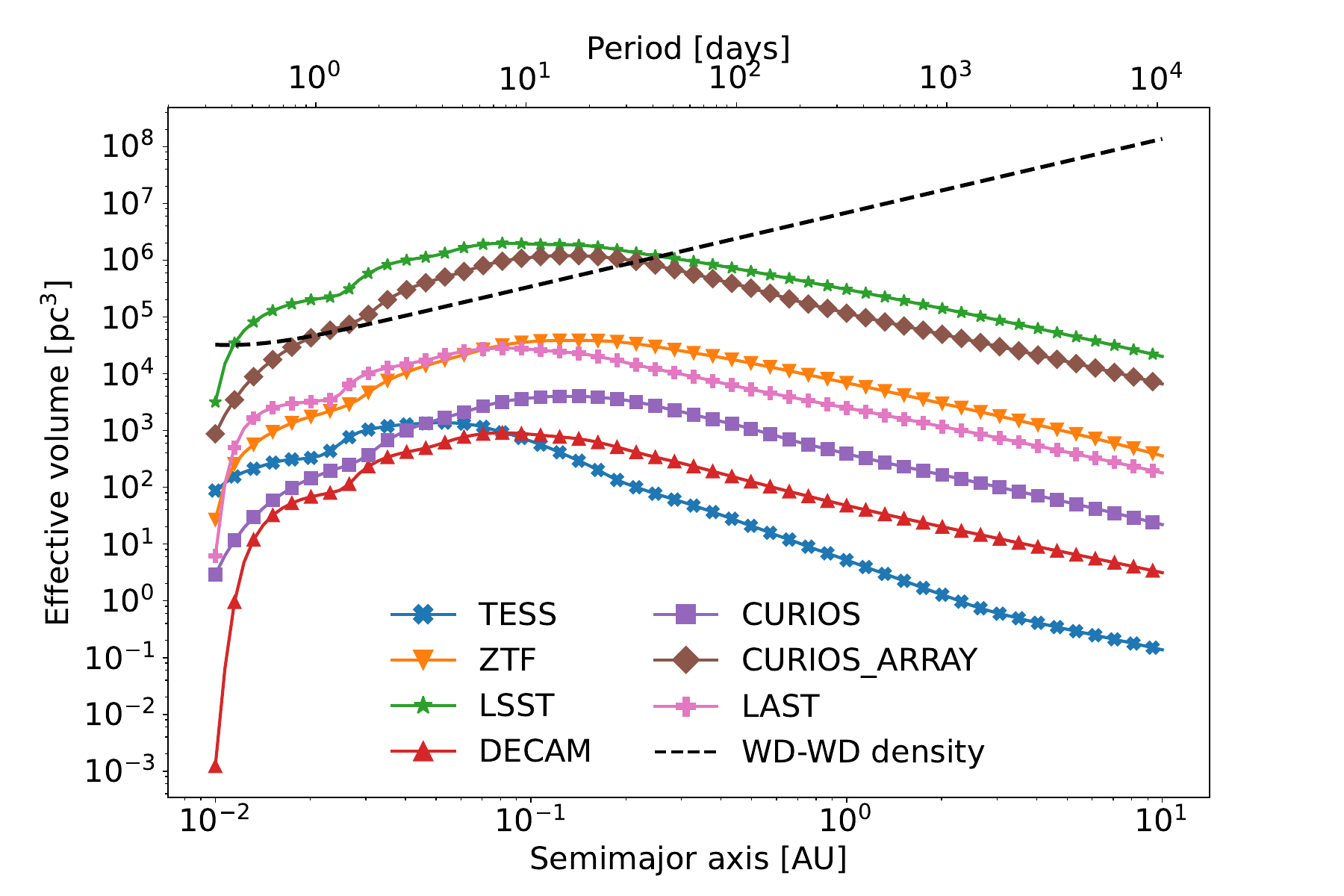}
    \caption{Effective volume for detecting WD-WD binaries using different surveys. 
    The dashed line represents an estimate for the space density 
    of double white dwarfs of one double WD (DWD) per 1818\,pc$^{3}$ 
    \citep{white_dwarfs_binary_separation_Maoz_2018}. 
    We can see the great depth and substantial field of view of LSST
    give it a clear advantage in detecting self-lensing flares, 
    with the (substantially less expensive) array of CuRIOS satellites
    getting very similar results based on their all-sky coverage. 
    Two other noteworthy surveys are ZTF and the-soon-to-be-operational LAST, 
    that may also provide some detections over their life-spans. 
    The single CuRIOS, TESS and DECam all have insufficient depth or coverage
    to detect meaningful number of WD-WD self-lensing pairs. 
    }
    \label{fig: effective volume wd}
\end{figure}

We present the effective volume for WD-WD pairs in Figure~\ref{fig: effective volume wd}. 
Each survey has slightly different effective volume curves, depending on the 
exact survey strategy, depth and cadence. 
We also show as a dashed line the estimated density of binary WDs in 
our Solar neighborhood \citep{white_dwarfs_binary_separation_Maoz_2018}. 
The normalization for WD-WD binaries is $5.5\times 10^{-4}$\,pc$^{-3}$
(or one system per 1818\,pc$^3$). 
From this plot it is evident that large surveys like LSST, 
or a large array of space telescopes like the full CuRIOS array,
have a volume coverage that easily surpasses the density of WD binaries. 
Other surveys like ZTF are not designed to detect self-lensing binaries,
that have a low duty cycle and often very faint flares, 
but still have a chance to see a few flares over their lifespan due 
to their large field of view. 
However, it should be noted that an array of smaller telescopes
like LAST (with much lower construction costs) 
could have similar performance in terms of finding self-lensing flares, 
and due to their multiple-image-per-visit strategy, stand a much better 
chance at identifying the flares. 
While TESS has good precision and cadence, it does not have the depth
required to see many WDs over the entire sky. 
The DDF survey of DECam, on the other hand, has substantial depth and good cadence, 
but the small field of view makes it unlikely that any lensing WDs 
exist in its footprint. 
It should be noted that even though the effective volume
curves of some surveys are below the model density, 
the integral over all semimajor axis bins can still yield
more than a single detection over the life of the survey (as discussed below). 

\begin{figure}
    \centering
    \includegraphics[width=1\linewidth]{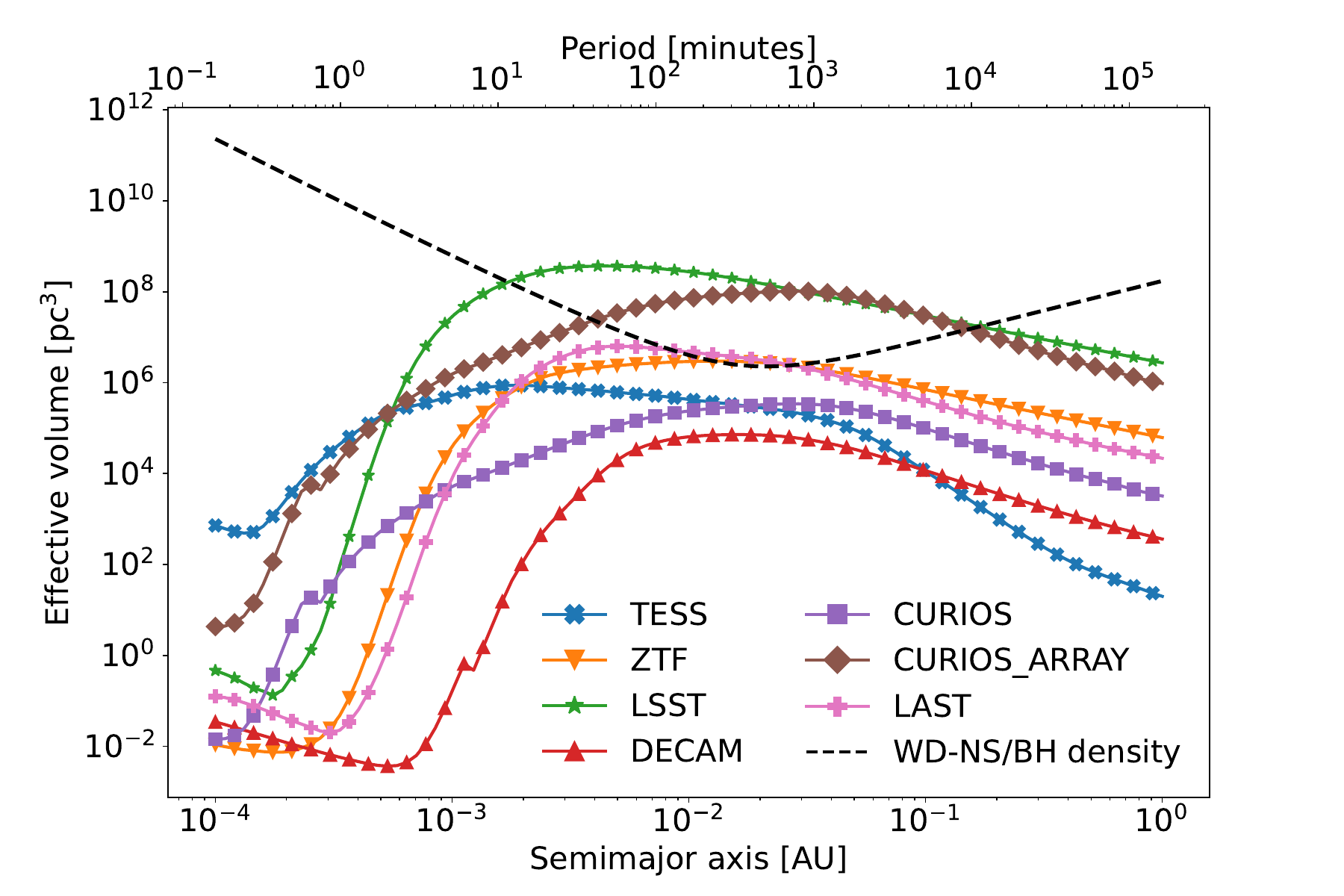}
    \caption{Effective volume for detecting WD-NS/BH binaries using different surveys. 
             The dashed line uses the same semimajor-axis distribution 
             as in Figure~\ref{fig: effective volume wd} 
             from \cite{white_dwarfs_binary_separation_Maoz_2018}, 
             adapted for higher-mass objects such as NSs and BHs. 
             The overall normalization of $10^{-5}$ per pc$^3$ 
             is arbitrarily chosen, as is the initial semimajor axis 
             power law index and the mass power law index (see text).
             Here, too, we see that LSST, with its superior depth, 
             and CuRIOS (full array), with its high cadence and sky coverage,
             have superior detection prospects relative to the other surveys. 
    }
    \label{fig: effective volume bh}
\end{figure}

For WDs in binaries with NSs or BHs we have only rough estimates for 
the intrinsic density of systems in the Solar neighborhood. 
While population synthesis models exist 
(e.g., \citealt{population_black_holes_Wiktorowicz_2019, populations_popsycle_Lam_2020}), 
we choose to simply present a rough estimate based on a reasonable density value, 
with the understanding that the total number of detections 
will scale linearly with the density normalization.
In Figure~\ref{fig: effective volume bh} we show the effective volume
compared to a distribution of NS/BH objects with the same semimajor axis 
distribution as in Figure~\ref{fig: effective volume wd}, 
but for more massive objects, with masses distributed 
according to a power law with an index of $-2.3$. 
This plot shows that, as before, a deep survey like LSST, 
or a shallow survey with all-sky coverage like CuRIOS (full array) 
will have dramatically better chances of seeing 
self-lensing flares than other surveys in our simulations. 
Note that the range of semimajor axes, and so the orbital periods, 
spans much shorter values in this figure. 

The corresponding flare durations of seconds to minutes 
may make it very hard to identify the flares are astrophysical, 
rather than instrumental or coming from some foreground effects, 
when they are seen in only one, isolated exposure. 
Furthermore, care should be taken when 
considering the semimajor axis distribution of massive compact objects.  
While we do not know the shape of the orbital distribution 
(or overall normalization) of WD-NS or WD-BH binaries, 
we can use this rough estimate to show what regions 
of the semimajor-axis parameter range are most likely to provide detections. 
Since the semimajor axis distribution is based on an initial power law, 
modified by gravitational wave emission \citep{white_dwarfs_binary_separation_Maoz_2018}, 
the overall shape should be similar to the WD-WD case, 
although we note that the initial power law index is unknown for 
NSs/BHs and is copied from the WD case as a placeholder, 
as is the arbitrary normalization of $10^{-5}$ systems per pc$^3$
(including WD-NS and WD-BH systems on a single mass spectrum 
with a single power law). 

To estimate the total number of detections in each survey we must integrate
the product of the effective volume with the probability to find each parameter:
\begin{equation}
    N_\text{total} = \int V(\Theta)P(\Theta) d\Theta, 
\end{equation}
where the effective volume V and the probability to find a system P
are both functions of the set of parameters $\Theta$
that include the masses, temperatures, inclination and semimajor-axis of each system. 
In the case of DWDs, the space density and distributions of mass, temperature 
and separation are relatively well known (e.g., \citealt{white_dwarfs_binary_separation_Maoz_2018}), 
so we use those estimates to get a rough number of detection in each survey.

For NSs and BHs in binaries with WDs, the numbers are essentially unknown, 
and can currently be estimated only using population synthesis codes. 
We will simplify the process by plugging in some range of reasonable distributions, 
to see how much the results change. 
The critical missing value, obviously, is the total number density of 
WD-NS or WD-BH pairs in the solar neighborhood. 
We will assume this number (over all masses and separations) is simply 
given as one system per $10^5$\,pc$^3$, 
and caution that the numbers we get based on this assumption
should be rescaled to more realistic numbers using, e.g., population synthesis codes. 
The results are summarized in Table~\ref{tab: number of detections}. 
The exact choice of power law index for the mass of NS/BH companions changes the 
resulting number of detections by up to $\pm20$\%, but given the uncertainties
in the overall normalization (the total number density of WD-NS and WD-BH pairs)
is probably not the main cause for uncertainty. 

\begin{deluxetable}{ccccc}

    \centering
    \tabletypesize{\small}
    \tablewidth{0pt}
    \tablecolumns{5}
    \tablecaption{Expected number of detections}
    \tablehead{
    \colhead{Survey} & 
    \colhead{WD-WD} & 
    \multicolumn{3}{c}{WD-NS/BH} \\
    \colhead{} & 
    \colhead{} & 
    \colhead{low} & 
    \colhead{mid} & 
    \colhead{high}}
    \startdata
    TESS & 0.3 & 2.2 & 2.5 & 2.8  \\
    ZTF & 3.3 & 18.0 & 21.3 & 24.7  \\
    LSST & 246.8 & 1314.8 & 1466.1 & 1614.5  \\
    DECAM & 0.1 & 0.4 & 0.5 & 0.6  \\
    CURIOS & 0.3 & 2.0 & 2.5 & 3.0  \\
    CURIOS\_ARRAY & 93.1 & 586.2 & 742.1 & 907.5  \\
    LAST & 3.7 & 23.6 & 26.9 & 30.3 
    \enddata
    
    \tablecomments{The low, mid and high columns correspond to 
                   power law indices for the NS/BH mass distribution.
                   In the simulations we used $-2.7$, $-2.3$ and $-2.0$
                   as the indices for the low, mid and high, respectively. 
                   The larger (less negative) index implies 
                   more BHs with higher mass, 
                   which increases the number of detections. 
                   The total number density for WD-WD pairs is 
                   taken to be one system in 1818\,pc$^3$
                   \citep{white_dwarfs_binary_separation_Maoz_2018}.
                   We assumed the total number density 
                   of all binaries of a WD with a NS/BH companion 
                   (with $M>1.5$\,M$_\odot$) are one system per $10^5$\,pc$^3$. 
                   This density is only given as a placeholder,
                   to be rescaled with the expected real number density from, 
                   e.g., population synthesis codes. }
    \label{tab: number of detections}

\end{deluxetable}


While some of the detection numbers in Table~\ref{tab: number of detections} may seem high, 
we caution that some of the systems we discuss will have a long period (of years)
and would be very hard to follow up. 
To get a rough estimate for how many systems we would actually be able to identify, 
we calculate the probability of each system to be identifiable in a dedicated followup campaign. 
As an example, we will assume a dedicated telescope will spend an average 2.4 hours each night 
following up a WD that had displayed a flare, over a whole year. 
The probability to see a flare during that time is:
\begin{equation}\label{eq: followup probability}
    P_\text{followup} = 1 - (1 - f)^N,
\end{equation}
where $f=0.1$ is the fraction of time the target is observed, 
and $N=\lfloor T_\text{followup} / T_\text{orbit}\rfloor$, 
with $T_\text{followup}=1$\,yr and $T_\text{orbit}$ being the orbital period. 
Note that systems with $T_\text{orbit}>T_\text{followup}$ will not be included
in the total effective volume and will not contribute to the total number of detections. 
Of course, followup could be distributed 
to multiple telescopes around the world, 
raising $f$ towards unity; 
On the other hand, many candidates would need to be followed up simultaneously, 
reducing the time a telescope can observe each target to a few minutes per night, 
such that $f$ could also be much smaller; 
and clearly, the limit of one year is arbitrarily chosen. 
These parameters are used to estimate how the total number of detections
would be limited by the requirement that systems are within reasonable 
possibility of a followup campaign. 

\begin{figure}
    \centering
    \includegraphics[width=1\linewidth]{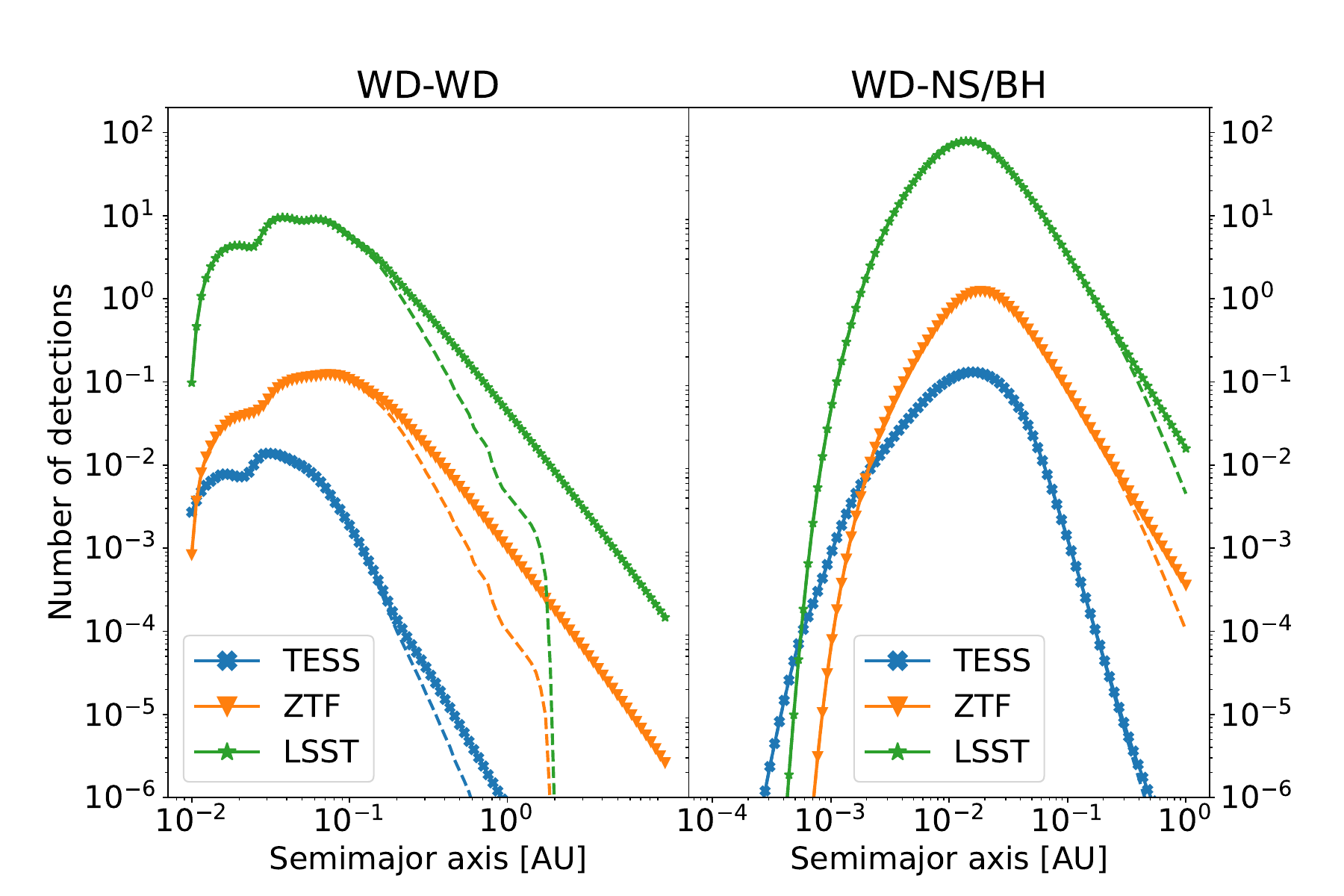}
    \caption{The number of WD-WD (left) and WD-NS/BH (right) 
             detections as a function of semimajor axis for 
             three representative surveys.
             The dashed lines represent the number of detected
             flares that could reasonably be followed up within 
             a year of 10\% observing fraction (see text). 
             The effect of a year of followup 
             are more substantial for the WD-WD case, 
             where the orbits are wider. 
             It should be noted that if the followup campaign 
             is shorter than a year 
             (e.g., due to funding or sun angle constraints)
             the effects would be felt at closer orbits. 
             }
    \label{fig: detection numbers}
    
\end{figure}

In Figure~\ref{fig: detection numbers} we show the number of detections 
(the product of the effective volume and the assumed number density of systems)
as a function of semimajor axis for WD-WD systems (left panel)
and for WD-NS/BH systems (right panel). 
Note the two panels use a different range of semimajor axis values. 
The overall normalization of the detection curves depends 
on the semimajor axis bin size, 
and should be integrated over all bins to get the total number of detections, 
as shown in Table~\ref{tab: number of detections}. 
However, the shape of these curves also help guide surveys
to look for each type of system at the optimal range of orbital periods for detection. 
Since the assumed distributions of compact binaries as a function of semimajor axis have 
a peak that partially overlaps with the peak in effective volume
(see Figures~\ref{fig: effective volume wd} and \ref{fig: effective volume bh}), 
the shape of the expected detections curves is qualitatively similar 
to the effective volume curves. 
The dashed lines in Figure~\ref{fig: detection numbers} show the number
of detections that can also be identified using the followup campaign 
described by Equation~\ref{eq: followup probability}. 
The WD-WD systems are more affected by this requirement, 
as they generally need wider orbits to be detectable. 
The chosen followup campaign parameters give
a gradual decline starting at orbits of a few weeks, 
and a sharp drop-off at one-year orbits. 
This does not affect the peak of the distribution
that occurs for most surveys around 10 days. 
However, shorter followup campaigns 
(due to, e.g., funding limits or sun-angle constraints) 
would shift this decline in numbers to shorter periods, 
and could impact the chances of detecting and identifying self-lensing flares.

\begin{figure}
    \centering
    \includegraphics[width=1\linewidth]{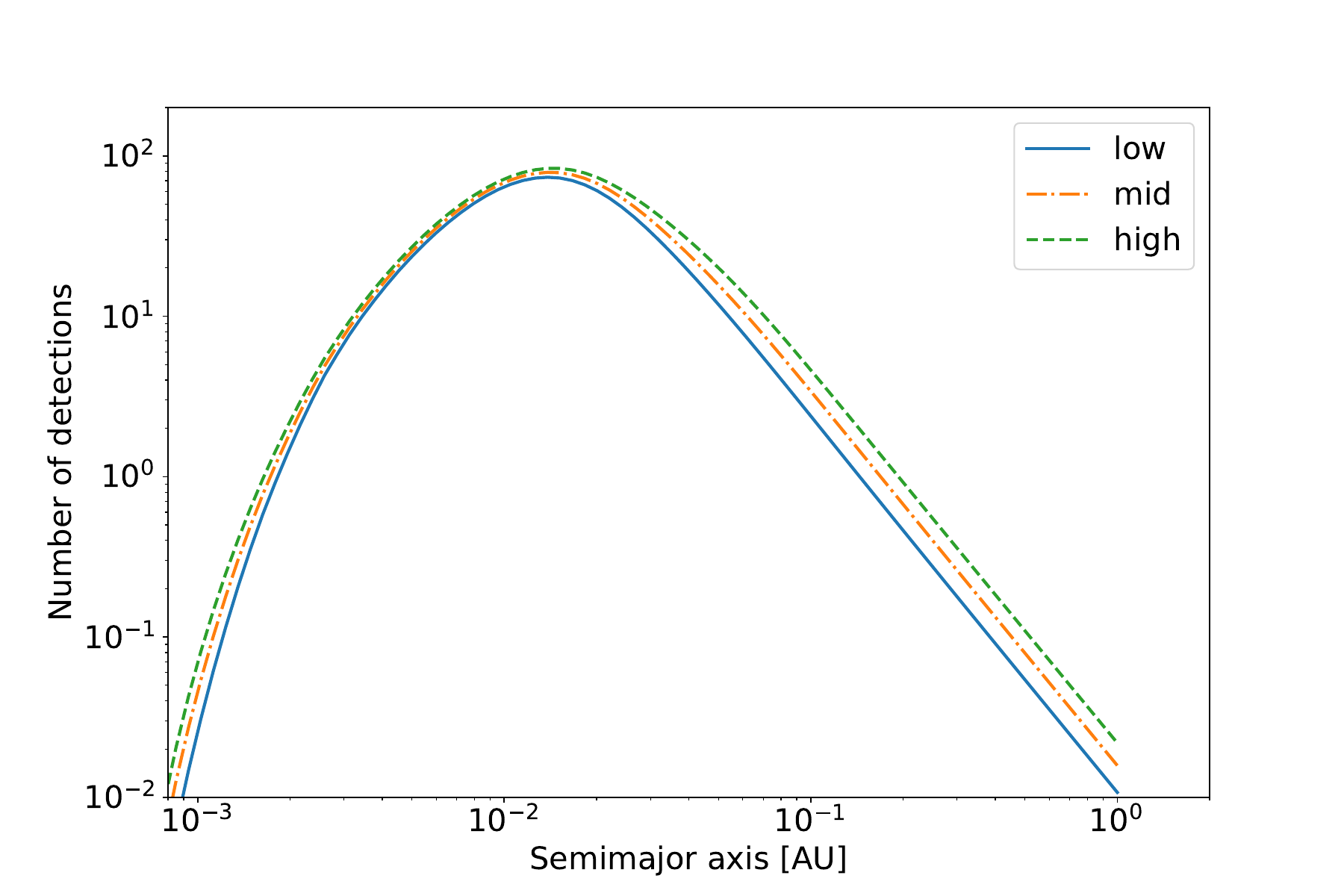}
    \caption{The number of WD-NS/BH detections as a function of semimajor 
             axis for the LSST survey, based on different choices of 
             mass power law index. 
             The total number of detections changes by up to a factor of 2, 
             but the overall shape of the curve does not change dramatically. 
             The index choices used in this work are $-2.7$, $-2.3$ and $-2.0$, 
             for the low, mid and high mass distributions, respectively.  
             }
    \label{fig: detections mass index}
    
\end{figure}

To get a sense of how that total number of detections would change
using different mass distribution models for NS/BH lenses, 
we integrate the effective volume multiplied by different probability 
functions for the mass of the lens in each binary system. 
Besides the model already presented in previous figures (using an index of $-2.3$), 
we also use a "low" and "high" indices of $-2.7$ and $-2.0$, respectively. 
The model with the less negative value (the ``high'' model) has more black holes
with a high mass, that are generally easier to detect. 
We show the number of WD-NS/BH detections expected for LSST, 
using different power law indices in Figure~\ref{fig: detections mass index}. 
We see that the numbers change by up to a factor of 2 
at some range of semimajor axis values, 
but we also note that when integrating all the detections 
we are left with a difference of about 20\% in either direction from 
the ``mid'' distribution (as seen in Table~\ref{tab: number of detections}). 

\begin{figure}
    \centering
    \includegraphics[width=1\linewidth]{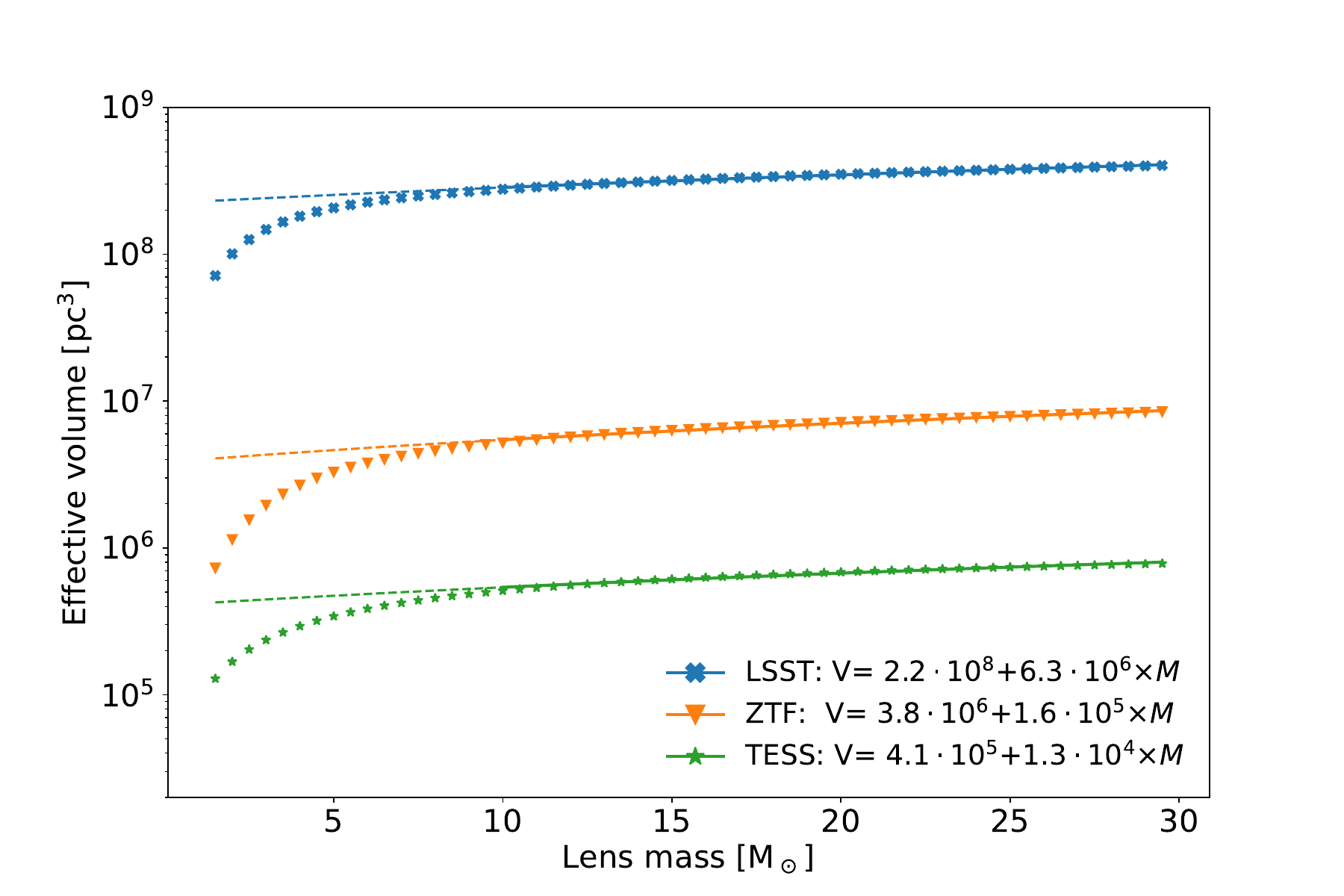}
    \caption{The effective volume of WD-NS/BH detections as a function of mass
             for three representative surveys. 
             The semimajor axis has been marginalized over using the same
             distribution shown as a dashed line in Figure~\ref{fig: effective volume bh}. 
             We fit a linear function to each curve, using only values of $M_L>10$\,M$_\odot$, 
             where the effective volume grows linearly with mass (solid curve). 
             The dashed curve shows the extrapolation of the linear function, 
             to regions where the effective volume grows faster with the lens mass
             than the fitted linear function. 
             }
    \label{fig: effective volume mass}
    
\end{figure}

To further explore the way our results would change under
arbitrary choices of mass distributions\footnote{ 
    E.g., a realistic mass distribution that has a different shape
    for NSs and for BHs, that would presumably arise 
    from different processes of stellar deaths that produce NSs vs.~BHs. }  
we show the effective volume of WD-NS/BH systems as a function 
of lens mass, this time marginalizing over the semimajor axis distribution
(we use the same distribution used above for modeling the WD-NS/BH population). 
We show the results, for three representative surveys, 
in Figure~\ref{fig: effective volume mass}.

For each survey we attempt to fit the effective volume curve 
using a linear function of the lens mass, $M_L$. 
At low masses, it seems the effective volume increases quickly on a curved slope, 
presumably due to the increased parameter space at closer orbits 
that becomes available as the mass increases and the flares become detectable at shorter orbits. 
At high mass values we expect that the Einstein radius will increase as $\sqrt{M_L}$ 
(see Equation~\ref{eq: einstein radius}, 
which contributes to increasing the geometric factor and the duty cycle. 
Together, the effective volume increases as a linear function of mass. 
To fit this behavior, we used only effective volume values where $M_L>10$\,M$_\odot$
to find the parameters of the linear function 
(the solid line represents the area used in the fit, 
while the dashed line shows the continuation of the line 
down to the region where the effective volume is non-linear with lens mass).

\section{Discussion}\label{sec: discussion}

Detecting WD-WD self-lensing binaries is difficult. 
The requirement for the lensing to overcome the physical eclipse
means that only pairs with relatively wide orbits can be detected. 
Eclipsing WD-WD binaries, on the other hand, with shorter orbits 
of minutes up to about an hour, are easier to detect due to their
higher geometric factor and duty cycle. 
Even though the intrinsic density of DWDs decreases with orbits shorter 
than the peak (around one day orbits), 
the prospects for detecting the minute to hour eclipsing binaries would be higher. 
The fact that with closer orbits the physical size of the eclipsing WD
does not shrink (as does the Einstein radius) means that the detection 
probability will scale with $a^{-2}$, instead of $a^{-1}$ that is typical 
of self lensing events. 

Indeed, cases of successful identification of WD-WD eclipsing binaries 
in archival ZTF data have been reported 
\citep{eclipsing_wds_shortest_orbit_Burdge_2019, eclipsing_wds_second_ZTF_Coughlin_2020, eclipsing_wds_second_shortest_Burdge_2020}. 
In these cases, the short orbital periods and associated high duty cycle
benefited from a folding analysis that combined the S/N from 
multiple eclipses that allowed detection of fainter signals than
what would be possible using individual eclipses. 
Considering that self-lensing flares would only be detectable, 
for WD binaries, with orbital periods of $\sim 1$ day, 
it is reasonable to expect an order of magnitude fewer 
detections of self-lensing flares in the entire ZTF archive. 
These longer orbits would also make it very unlikely to see
more than one flare per system, so a search based on folding 
all epochs in the archive over different test periods would
most likely not improve the detection prospects. 

Our results for ZTF (in Table~\ref{tab: number of detections}) 
estimate that there should be around three flares detected from 
three separate WDs during five years of ZTF operations. 
This is roughly consistent with the number of eclipsing WDs 
in the ZTF archive. 
As we would expect an order of magnitude less self-lensing DWDs
to be detectable in the same dataset, 
the estimates given by our simulations are perhaps somewhat optimistic. 
Searching for such flares in archival ZTF data is underway. 

WDs in close binaries with NSs or BHs would not have any detectable eclipses, 
and can be seen only by light-curve variability caused by elliptical modulations, 
Doppler beaming or self lensing \citep{black_hole_binaries_TESS_Masuda_2019, black_hole_binaries_photometric_Chawla_2023}.
The shorter periods and higher geometric factors and duty cycles relevant 
for detection of these more massive objects should yield more detections, 
provided the underlying number of systems is high enough. 
Searches already conducted for eclipsing DWDs in ZTF data 
had not led to the discovery of any WD-NS or WD-BH self-lensing flares.
This may be an indication that the number density of such systems is lower 
(by an order of magnitude or more)
than the placeholder value of $10^{-5}$\,pc$^{-3}$ we used to produce 
the values in Table~\ref{tab: number of detections}. 

Considering, instead, the detection prospects for a survey such as TESS, 
we may start by assessing the number of available WDs that are observable
with sufficient photometric precision. 
Using the WD catalog assembled by \cite{white_dwarfs_in_Gaia_DR3_Fusillo_2021}, 
we find a total of 4125 WDs brighter than $G<16$, 
or $\approx 33000$ WDs brighter than $G<18$ 
(magnitudes given by the Gaia $G$ band). 
Since DWDs constitute about 10\% of the total WD population, 
and since the geometric factor for self-lensing systems with 
the appropriate $\sim 1$ day orbit is $\approx 1/1000$, 
we would expect less than one detection in the $G<16$ census of WDs
that may be bright enough to be successfully monitored by TESS. 
This back-of-the-envelope calculation is consistent with our simulation results. 
However, a search for faint, symmetrical flares using a matched-filter approach 
\citep{matched_filter_Turin_1960}, 
perhaps also using folding on a range of trial periods may be warranted, 
to verify that an unknown population of DWDs is not hiding in the data. 

How would our results change if we considered main sequence stars
instead of WDs as the sources that are being lensed? 
While a detailed analysis with a full simulation set is reserved
to future work (and has already been considered elsewhere
\citealt{self_lensing_large_sources_Agol_2003, self_lensing_main_sequence_sources_Rahvar_2011, self_lensing_MS_NS_BH_Wiktorowicz_2021}), 
we can make some general remarks based on scaling the current results. 

A main sequence star like our sun, is about 100 times the radius of a typical WD. 
Even if we consider smaller stars, the sources will be larger and the flares 
more diluted, unless the orbits are also substantially larger. 
To preserve the same ratio of the Einstein radius to the source radius, 
for a star $X$ times larger than the typical WD, 
would require an orbit with a semimajor axis that is $X^2$ larger
(see Equation~\ref{eq: einstein radius}). 
This leads to an increase of orbital period of $X^3$. 
Thus, a one-day DWD binary will have a similar flare magnitude
as a 1000-day M-dwarf-WD binary, assuming the M-dwarf is just 
10 times larger than the WD. 
The overall phase space (geometric factor times the duty cycle) 
scales as $a^{-1}$ so the number of detectable flares would shrink
as $X^{-2}$ when considering these larger sources. 
It should be noted that surveys with substantially better
photometric precision can be sensitive to shorter orbit
MS-WD binaries, as indeed was the case for five different
systems found in the Kepler footprint 
\citep{self_lensing_MS_WD_first_system_Kruse_Agol_2014, self_lensing_MS_WD_three_systems_Kawahara_2018, self_lensing_MS_WD_wide_orbit_Masuda_2019}. 

The reduced probability to see a flaring MS-WD system with a wider orbit
does not mean that such systems will be harder to find. 
WDs are a small population and are much fainter than MS stars. 
Under the simplifying assumption that the total number density of DWD systems
is the same as that of our hypothetical MS-WD systems where the MS star's 
radius is $X$ times the radius of the WD, 
the distance at which our MS systems are visible in space
grows also as $X$ (under the wrong assumption of equal temperature). 
Under all these overly simplistic assumptions, 
the factor of $X^3$ increase in MS-WD systems caused by the larger
accessible space volume compensates for the lower geometric factor and duty cycle
of the larger orbits (which scales as $X^{-2}$ as mentioned above). 
Obviously, the exact distribution of stellar sizes and temperatures
should be used to calculate the expected number of detections, 
using similar methods to those presented in this work. 
Even with such simple assumptions, 
it is plausible that there would be no shortage of wide-orbit, 
MS-WD systems that can be detected with the same $\sim 1$\% precision 
characteristic of most ground based telescopes. 

However, the fact that such events would have a repeat period
of $\approx 1000$ days even for $X\approx 10$ means that any such 
systems that are detected would be very difficult to follow up. 
A survey like LSST, which stands a good chance of detecting many such flares
in a single visit (of one 30\,s exposure or two 15\,s exposures) 
over the survey lifespan, 
would not have any simple way to differentiate the flare from a 
stellar eruption or other types of artefacts like satellites or cosmic rays. 

The results for WD-NS and WD-BH systems are heavily model-dependent. 
Since there are no direct measurements of populations of WDs in binaries 
with heavier compact objects, we are left with population synthesis codes. 
While we do not expect the (probably optimistic) numbers in Table~\ref{tab: number of detections}
to reflect the actual numbers that would be detected, 
the converse may become true: 
with continued observations and sufficient, dedicated photometric followup campaigns,
we stand a good chance of constraining this unknown population of 
massive, compact objects based on detection (or lack thereof) of 
self-lensing flares. 

The parameter space for detecting these more massive binaries
is not constrained by eclipses, 
so that much shorter orbits are still viable targets for ground based surveys. 
For the relatively wider orbits required for detecting MS stars 
in binaries with NSs or BHs the parameter space is also still viable.
Where a WD-BH system would be most likely detected at minutes-hours orbits
(see Figure~\ref{fig: mag vs period}) 
replacing the WD with a small MS star increases 
the relevant orbits to the few-day period range, 
which is still accessible to followup with relatively minor 
investment of resources. 

In all the cases discussed above, 
whether the source is a WD or a MS star, 
and in all but the shortest orbits where BH companions
could conceivably be detected, 
the expected flares would present as a symmetric 
signal over minutes or hours. 
This raises the importance of surveys that
employ long visits of consecutive exposures on the same field. 
Not only would such strategy help rule out asteroids, satellites and cosmic rays,
when multiple exposures show the signal appear and fade away, 
but having additional information on the shape of the light-curve
would be pivotal in identifying the best candidates for self lensing. 

Another way to distinguish self-lensing flares would be 
by simultaneously observing in multiple filters. 
Up to corrections due to limb-darkening, 
self-lensing flares will be achromatic, 
while stellar flares are expected to be blue. 

These observing strategies are especially important 
for detecting flares from MS sources, 
particularly from smaller, late type stars, 
as they will inevitably suffer from a background of 
stellar flares that could have similar time-scales to self-lensing flares, 
but with very different spectra and morphology \citep{stellar_flare_foreground_Kulkarni_Rau_2006}.

\section{Conclusions}\label{sec: conclusions}

We present a simulation tool to efficiently reproduce light-curves
of self-lensing flares in a wide range of system parameters.\footnote{\url{https://github.com/guynir42/self_lens}} 
In addition, the tools we developed include methods for simulating
sky surveys with various strategies, allowing a calculation of 
the expected number of detections for different populations of self-lensing binaries. 

Using these tools we map the parameter space of self-lensing system
and provide some estimates for ranges of parameters that are most relevant
for detecting either WD-WD systems or WD-NS and WD-BH systems, 
that are qualitatively different as the former is affected by eclipses 
while the latter is more massive and has negligible eclipses. 
A major result of this work is given in Figure~\ref{fig: mag vs period}, 
which is a map for the complicated landscape describing the detection prospects
of different systems. 
The interplay of photometric precision, orbital period and duty cycle 
(which is also a good estimate for the geometric factor)
are mapped out for future surveys searching for self-lensing flares. 

We presented a few example surveys that show how different strategies
could be more or less effective in detecting self-lensing flares. 
Notably, the LSST survey dominates the detection potential due to its
formidable combination of depth, field of view and cadence. 
However, we caution that even with the potential to detect hundreds of flares, 
it is not certain that such a survey will be able to differentiate 
self-lensing flares from other fast transients and artefacts. 
On the contrary, an array of small satellites dedicated to high-precision 
photometry of a large fraction of the sky (e.g., the planned CuRIOS array)
can provide a similar yield of detected flares, 
but with a much denser cadence that would make it much easier to 
identify the nature of the flares. 
The lower depth of such a mission also allows cross-matching to other catalogs
and identifying the sources of flares as being small stars or WDs 
(e.g., using Gaia, as in the case for identifying WDs; \citealt{white_dwarfs_in_Gaia_DR3_Fusillo_2021}). 

Another survey we considered was ZTF. 
Despite being designed to detect slower transients on timescales 
of days and weeks, ZTF still has some potential for detecting self-lensing flares. 
Efforts to scan archival data for flare-like events from WD source is underway. 
However, we caution that the lower cadence will likely not allow 
detection of the same source with multiple exposures, 
so here, too, identification of self lensing would be difficult. 
With a handful of eclipsing DWD systems already detected by ZTF, 
a comparison of the detection prospects for self-lensing flares
is consistent with having a between zero to a few detections 
over the lifespan of the survey. 
A survey with similar grasp to ZTF that is currently under commissioning is LAST, 
which has similar detection statistics but has the advantage
of making visits of multiple exposures in a single field, 
which, again, is instrumental in identifying candidates for self lensing. 

We expect fewer than one DWD self-lensing flare to be 
detected in TESS data. The limiting factor in this case is the 
shallow depth that is accessible to the survey, 
yielding a relatively small number of WDs to monitor. 

Finally, we discuss the detection prospects of WDs
in binaries with NSs and BHs. 
Our results are fairly robust in terms of the 
range of orbital periods that are most likely to be detectable, 
and refer again to the map of the parameter space in Figure~\ref{fig: mag vs period}. 
We caution, however, that the underlying population of such binaries is 
essentially unknown, and even with sophisticated population synthesis codes, 
the channels to produce systems with such a large mass-disparity between 
the binary components are very much model-dependent. 
And so, admitting our lack of knowledge of the number density 
(or even existence) of such systems, 
we provide very rough estimates for the detection prospects in different surveys. 
On the other hand, (non-) detection of NS/BH self-lensing flares 
(with WD or MS source stars) could be a promising channel for constraining
populations of which we know very little. 
Surveys such as CuRIOS (and to a lesser degree, LAST) 
have a suitable visit strategies for detecting and identifying 
short-lived flares. 
We argue that an informed search for flares of the appropriate
shape, length and periodicity would be a valuable endeavor, 
that could shed light on an otherwise unknown, dark population.

\section{Data Availability}

The code used to generate all simulations, tables and figures
can be found at \url{https://github.com/guynir42/self_lens}\footnote{
The precise commit used in this work is at \url{https://github.com/guynir42/self_lens/commit/3977424c6e2944603e7330ebdc2339ebfe5cd95e}. 
}.
The source matrices used to speed up simulation times are included
in the above code repository and are also available as a Zenodo dataset
at \url{https://zenodo.org/record/8418767}. 
They can be regenerated locally using the \texttt{scripts} folder
using a different set of parameters, if needed. 
The effective volume datasets used as the basis for the results in this work 
can be simulated using the given code repository, 
but are also made available through two Zenodo datasets, 
one for DWD binaries: \url{https://zenodo.org/record/8340555};
and one for WDs with NS/BH companions: \url{https://zenodo.org/record/8340930}. 
These datasets are usually downloaded automatically by the \texttt{self lens}
package when running the test suite on a local machine. 

\section{Acknowledgements}

JSB and GN were partially supported by the Gordon and Betty Moore Foundation. 
We thank N.~Abrams, H.~Gulick and J.~Lu for fruitful discussions of 
WD-BH populations and the upcoming CuRIOS mission. 

\bibliographystyle{aasjournal}
\bibliography{self_lensing,microlensing,surveys,white_dwarfs,populations,algorithms,other_refs,eclipsing_wds}

\end{document}